\documentclass{aa}  
\usepackage{graphicx}
\usepackage{txfonts}
\usepackage{natbib}
\usepackage{url}
\def\fic{x_{\rm ic}}
\def\vinf{v_{\infty}}
\begin{document}
   \title{Mass loss from inhomogeneous hot star winds}

   \subtitle{I. Resonance line formation in 2D models}
   \author{J.O. Sundqvist\inst{1}\and
           J. Puls\inst{1}\and 
           A. Feldmeier\inst{2}
           }
          
   \institute{Universit\"atssternwarte M\"unchen, Scheinerstr. 1, 81679 M\"unchen, Germany\\
              \email{jon@usm.uni-muenchen.de}\and Institut f\"ur Physik und Astronomie,
              Karl-Liebknecht-Strasse 24/25, 14476 Potsdam-Golm, Germany}
   \date{Received 7 July 2009 / Accepted 17 November 2009}

 
\abstract
       { The mass-loss rate is a key parameter of hot, massive stars.  
       Small-scale inhomogeneities (clumping) in the winds of these stars
       are conventionally included in spectral analyses by assuming
       optically thin clumps, a void inter-clump medium, and a smooth
       velocity field.  To reconcile investigations of different diagnostics
       (in particular, unsaturated UV resonance lines vs. $\rm
       H_{\alpha}$/radio emission) within such models, a highly clumped wind
       with very low mass-loss rates needs to be invoked, where the
       resonance lines seem to indicate rates an order of magnitude (or even
       more) lower than previously accepted values.  If found to be realistic,
       this would challenge the radiative line-driven wind theory and have
       dramatic consequences for the evolution of massive stars. }
       { We investigate basic properties of the formation of resonance lines
       in small-scale inhomogeneous hot star winds with non-monotonic
       velocity fields.}
       { We study inhomogeneous wind structures by means of 2D stochastic and
       pseudo-2D radiation-hydrodynamic wind models, constructed by
       assembling 1D snapshots in radially independent slices. A Monte-Carlo
       radiative transfer code, which treats the resonance line formation in an
       axially symmetric spherical wind (without resorting to the Sobolev
       approximation), is presented and used to produce synthetic line
       spectra.}
       { The optically thin clumping limit is
       only valid for very weak lines. The detailed density structure,
       the inter-clump medium, and the non-monotonic velocity field
       are all important for the line formation. We confirm previous
       findings that radiation-hydrodynamic wind models reproduce
       observed characteristics of strong lines (e.g., the black
       troughs) without applying the highly supersonic
       `microturbulence' needed in smooth models.  For intermediate
       strong lines, the velocity spans of the clumps are of central
       importance. Current radiation-hydrodynamic models predict spans
       that are too large to reproduce observed profiles unless a very
       low mass-loss rate is invoked. By simulating lower spans in 2D
       stochastic models, the profile strengths become drastically
       reduced, and are consistent with higher mass-loss rates. To
       simultaneously meet the constraints from strong lines, the
       inter-clump medium must be non-void. A first comparison to the
       observed Phosphorus V doublet in the O6 supergiant
       $\lambda$~Cep confirms that line profiles calculated from a
       stochastic 2D model reproduce observations with a mass-loss
       rate approximately ten times higher than that derived from the
       same lines but assuming optically thin clumping.  Tentatively
       this may resolve discrepancies between theoretical predictions,
       evolutionary constraints, and recent derived mass-loss rates,
       and suggests a re-investigation of the clump structure
       predicted by current radiation-hydrodynamic models. }
  {}

   \maketitle
%

\section{Introduction}
\label{Introduction}

Mass loss through supersonic stellar winds is pivotal for the physical
understanding of hot, massive stars and their surroundings. A change of only
a factor of two in the mass-loss rate has a dramatic effect on massive star
evolution \citep{Meynet94}.  Winds from these stars are described by
the line-driven wind theory \citep{Castor75,Pauldrach86}, which
traditionally assumes the wind to be stationary, spherically symmetric, and
homogeneous.  Despite this theory's apparent success
\citep[e.g.,][]{Vink00}, evidence for an inhomogeneous and time-dependent
wind has over the past years accumulated, recently summarized in the
proceedings from the workshop `Clumping in hot star winds' \citep{Hamann08}
and in a general review of mass loss from hot, massive stars \citep{Puls08}.

That line-driven winds should be intrinsically unstable was already
pointed out by \citet{Lucy70}, and was later confirmed first by
linear stability analyses and then by direct, radiation-hydrodynamic
modeling of the time-dependent wind
\citep[e.g.,][]{Owocki84,Owocki88,Feldmeier95,Dessart05}, where the
line-driven (or line-deshadowing) instability causes a small-scale,
inhomogeneous wind in both density and velocity.

\textit{Direct observational} evidence of a small-scale, clumped stellar
wind has, for O-stars, so far only been given for two objects, $\zeta$ Pup
and HD\,93129A \citep{Eversberg98,Lepine08}.  Much \textit{indirect}
evidence, however, has arisen from quantitative spectroscopy, where the
standard way of deriving mass-loss rates from observations nowadays is via
line-blanketed, non-LTE (LTE: local thermodynamic equilibrium) model
atmospheres that include a treatment of both the photosphere and the wind.
Wind clumping has been included in such codes (e.g., CMFGEN
\citep{Hillier98}, PoWR \citep{Grafener02}, FASTWIND \citep{Puls05}) by
assuming statistically distributed \textit{optically thin} density clumps
and a void inter-clump medium, while keeping the smooth velocity law. The
major result from this methodology is that any mass-loss rate derived from
smooth models and density-squared diagnostics ($\rm H_{\alpha}$, infra-red
and radio emission) needs to be scaled down by the square root of the
clumping factor (which describes the over density of the clumps as compared
to the mean density, see Sect.~\ref{wind_stoch}). For example,
\citet{Crowther02}, \citet{Bouret03}, and \citet{Bouret05} have concluded
that a reduction of `smooth' mass-loss rates by factors $3 \dots 7$ might be
necessary. Furthermore, from a combined optical/IR/radio analysis of a
sample of Galactic O-giants/supergiants, \citet{Puls06} derived upper limits
on observed rates that were factors of $2 \dots 3$ lower than previous $\rm
H_{\alpha}$ estimates based on a smooth wind.

On the other hand, the strength of UV resonance lines (`P Cygni lines') in
hot star winds depends linearly on the density and is therefore not believed
to be directly affected by optically thin clumping. By using the Sobolev
with exact integration technique (SEI; cf.~\citeauthor{Lamers87} 1987) on
the unsaturated Phosphorus V (PV) lines, \citet{Fullerton06} for a large
number of Galactic O-stars derived rates that were factors of $10 \dots 100$
lower than corresponding smooth $\rm H_{\alpha}$/radio values (provided PV
is the dominant ion in spectral classes O4 to O7). Such large revisions
would conflict with the radiative line-driven wind theory and have dramatic
consequences for the evolution of, and the feedback from, massive stars
\citep[cf.][]{Smith06,Hirschi08}.  Indeed, a puzzling picture has emerged,
and it appears necessary to ask whether the present treatment of wind
clumping is sufficient. Particularly the assumptions of optically thin
clumps, a void inter-clump medium, and a smooth velocity field may not be
adequate to infer proper rates under certain conditions.

\paragraph{Optically thin vs. optically thick clumps.} \citet{Oskinova07}
used a porosity formalism \citep{Feldmeier03,Owocki04} to scale the opacity
from smooth models and investigate impacts from \textit{optically thick}
clumps on the line profiles of $\zeta$ Pup. Due to a reduction in the
effective opacity, the authors were able to reproduce the PV lines without
relying on a (very) low mass-loss rate, while simultaneously fitting the
optically thin $\rm H_{\alpha}$ line. This formalism, however, was
criticized by \citet{Owocki08} who argued that the original porosity concept
had been developed for continuum processes, and that line transitions rather
should depend on the non-monotonic velocity field seen in hydrodynamic
simulations. Proposing a simplified analytic description to account for this
velocity-porosity, or `vorosity', he showed how also this effect may reduce
the effective opacity.

In this first paper we attempt to clarify the most important concepts
by conducting a detailed investigation on the synthesis of UV
resonance lines from inhomogeneous two-dimensional (2D) winds. We
create both pseudo-2D, radiation-hydrodynamic wind models and 2D,
stochastic wind models, and produce synthetic line profiles via
Monte-Carlo radiative transfer calculations.  We account for and
analyze the effects from a wind clumped in \textit{both} density and
velocity as well as the effects from a non-void inter-clump
medium. Especially we focus on lines with intermediate line strengths,
comparing the behavior of these lines with the behavior of both
optically thin lines and saturated lines.  Follow-up studies will
include a treatment of emission lines (e.g., $\rm H_{\alpha}$) and an
extension to 3D, and the development of simplified approaches to
incorporate effects into non-LTE models.

In Sect.~\ref{wind} we describe the wind models and in Sect.~\ref{rt}
the Monte-Carlo radiative transfer code.  First results from 2D
inhomogeneous winds are presented in Sect.~\ref{2d}, and an extensive
parameter study is carried out in Sect.~\ref{ps}.  We discuss some
aspects of the interpretations of these results and perform a first
comparison to observations in Sect.~\ref{Discussion}, and summarize
our findings and outline future work in Sect.~\ref{Conclusions}.

\section{Wind models}
\label{wind}

\begin{figure*}
  \begin{minipage}{8.8cm}
    \resizebox{\hsize}{!}
              {\includegraphics[angle=90]{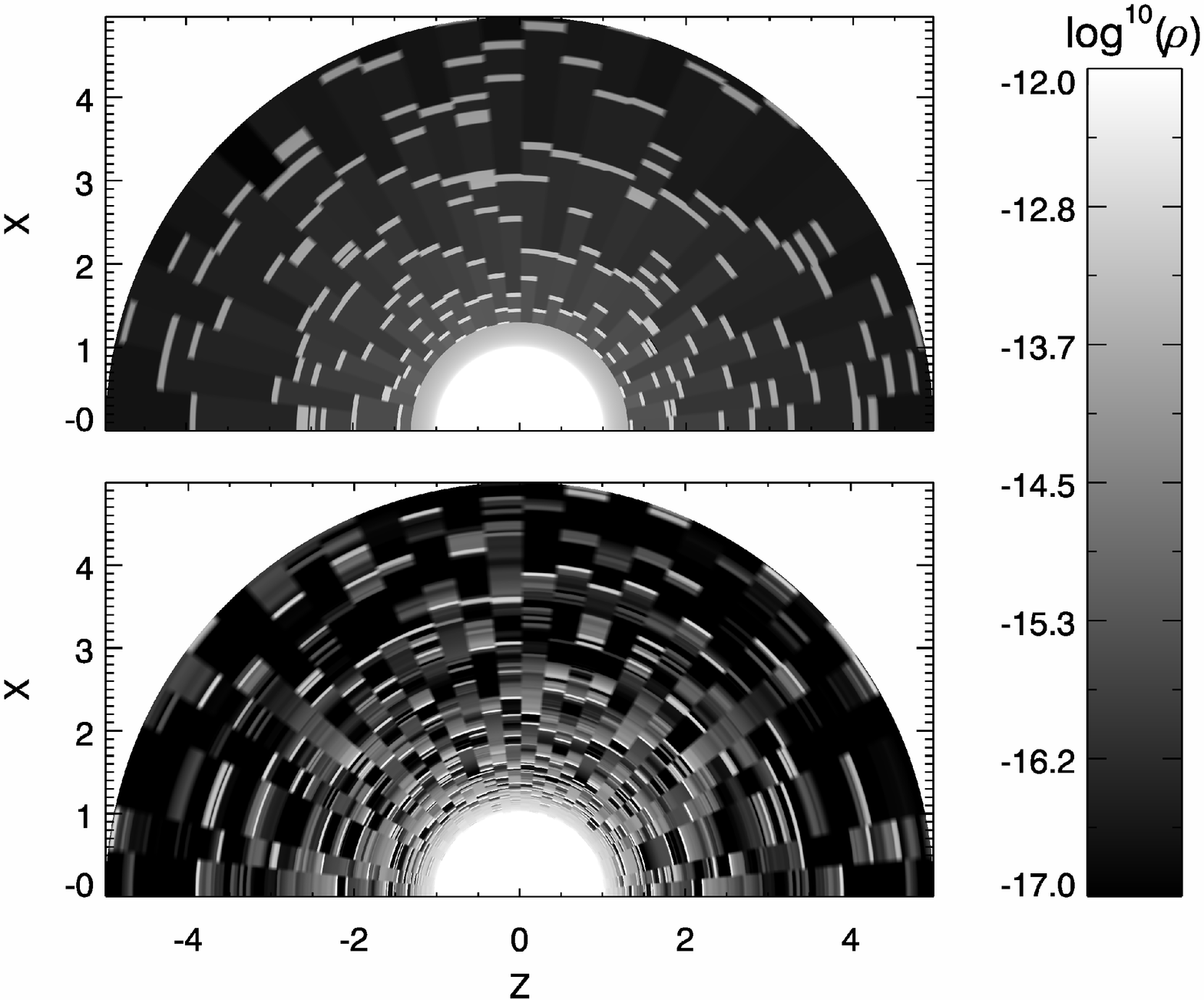}}
  \end{minipage}
  \begin{minipage}{8.8cm}
    \resizebox{\hsize}{!}
              {\includegraphics[angle=90]{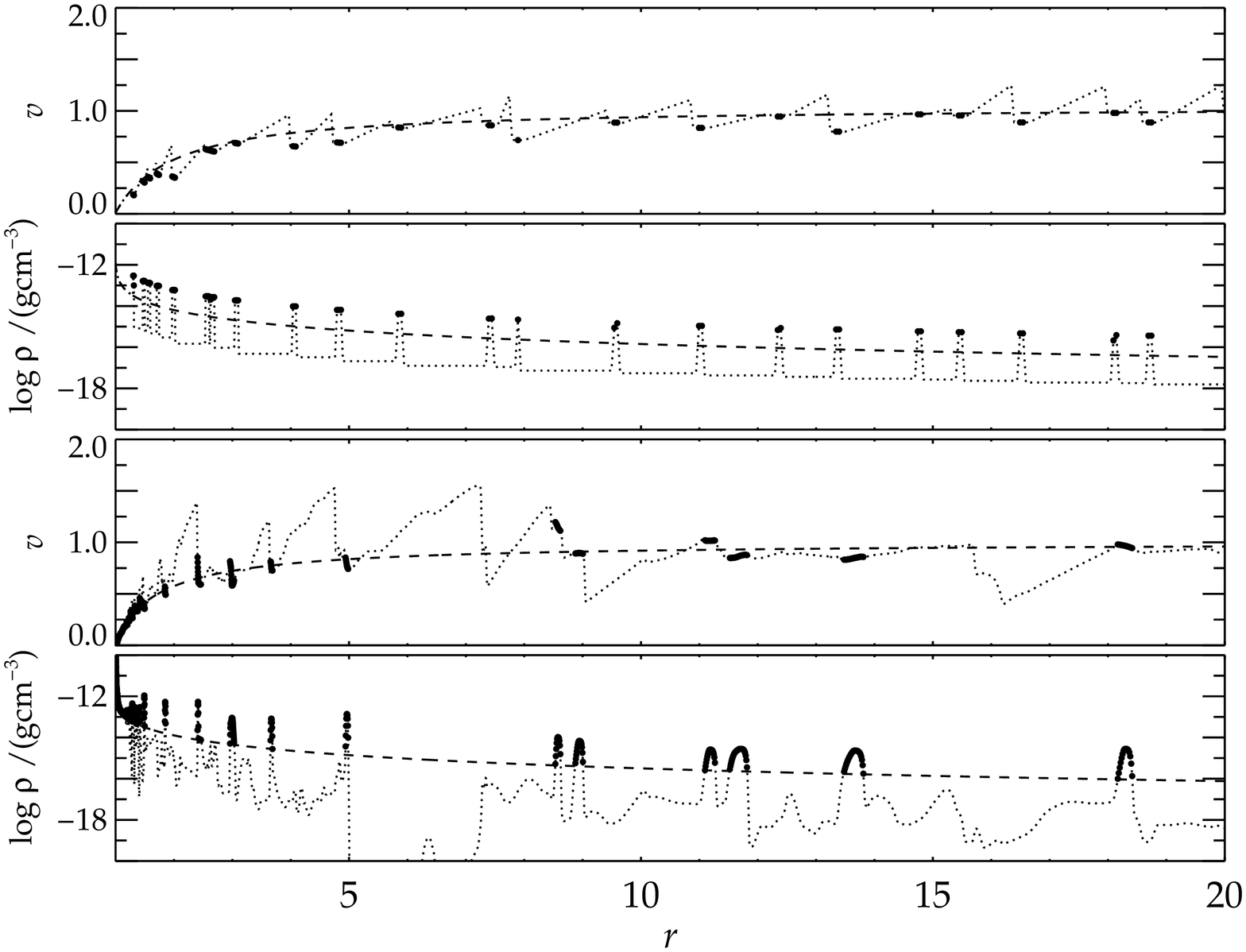}}
  \end{minipage}
  \caption{\textit{Left panel:} Density contour plots of one stochastic 
    (upper plot) and one RH (FPP, lower plot) model. The Cartesian 
    coordinate $Z$ is on the abscissa and $X$ is on the ordinate.
    \textit{Right panel:} Density and velocity structures of one slice 
    in one stochastic (upper) and one RH (FPP, lower) model. 
    Over densities are marked with filled dots.
    For model parameters and details, see Sect.~\ref{wind_stoch}.} 
  \label{Fig:contours}
\end{figure*}

For wind models, we use customary spherical coordinates $(r,\Theta,\Phi)$
with $r$ the radial coordinate, $\Theta$ the polar angle, and $\Phi$ the
azimuthal angle. We assume spherical symmetry in 1D models and symmetry in
$\Phi$ in 2D models.  In all 2D models $\Theta$ is sliced into $N_{\Theta}$
equally sized slices, giving a lateral scale of coherence (or an opening
angle) $180 / N_{\Theta}$ degrees. This 2D approximation is discussed in
Sect.~\ref{3d}.  Below we describe the model types primarily used in the
present analysis; two are of stochastic nature and two are of
radiation-hydrodynamic nature.

\subsection{Radiation-hydrodynamic wind models}
\label{wind_rh} We use the time-dependent, radiation-hydrodynamic (hereafter
RH) wind models from \citet[][hereafter `POF']{Puls93}, calculated by
S.~Owocki, and from \citet[][hereafter `FPP']{Feldmeier97}, and the reader
is referred to these papers for details.  Here we summarize a few important
aspects.  POF assume a 1D, spherically symmetric outflow, and circumvent a
detailed treatment of the wind energy equation by assuming an isothermal
flow.  Perturbations are triggered by photospheric sound waves.  The wind
consists of 800 radial points, extending to roughly 5 stellar radii. FPP
also assume a 1D, spherically symmetric outflow, but include a treatment of
the energy equation.  Perturbations are triggered either by photospheric
sound waves or by Langevin perturbations that mimic photospheric turbulence.
The wind consists of 4000 radial points, extending to roughly 30 stellar
radii.  Tests have shown that the FPP winds yield similar results for both
flavors of perturbations, and, for simplicity, we therefore use only the
results of the turbulence model.

Due to the computational cost of obtaining the line force, only initial
attempts to 2D RH simulations have been carried out
\citep{Dessart03,Dessart05}.  These authors first used a strictly radial
line force, yielding a complete lateral incoherent structure due to
Rayleigh-Taylor or thin-shell instabilities, and in the follow-up study uses
a restricted 3-ray approach to approximate the lateral line drag, yielding a
larger lateral coherence but lacking quantitative results. Therefore, and
because of the general dominance of the radial component in the radiative
driving, we create fragmented 2D wind models from our 1D RH ones by
assembling snapshots in the $\Theta$ direction, assuming independence
between each slice consisting of a pure radial flow.  After the polar angle
has been sliced into $N_{\Theta}$ equally sized slices, one random snapshot
is selected to represent each slice.  This method for creating more-D models
from 1D ones is essentially the same as the `patch method' used by
\citet{Dessart02}, when synthesizing emission lines for Wolf-Rayet stars,
and the method used by, e.g., \citet{Oskinova04}, when synthesizing X-ray
line emission from stochastic wind models. Fig.~\ref{Fig:contours} displays
typical velocity and density structures from this type of 2D model.
   
\subsection{Stochastic wind models}
\label{wind_stoch}

We also study clumpy wind structures created by means of distorting a
smooth, stationary, and spherically symmetric wind via stochastic
procedures. This allows us to investigate the impacts from, and to set
constraints on, different key parameters without being limited by the
values predicted by the RH simulations. For the underlying smooth
winds we adopt a standard $\beta$ velocity law
$v_\beta(r)=(1-b/r)^{\beta}$. Here and throughout the paper, we
measure {\it all} velocities in units of the terminal velocity,
$\vinf$, and {\it all} distances and length scales in units of the
stellar radius, $R_{\star}$. $b$ is given by $v(r=1)=v_{\rm min}$, the
velocity at the base of the wind. $v_{\rm min}=0.01$ is assumed,
roughly corresponding to the sound speed. For a given $\dot{M}$, the
homogeneous density structure then follows directly from the equation
of continuity. We choose $\beta=1$, which is appropriate for a
standard O-star wind and allows us to derive simple analytic
expressions for wind masses and flight times.
      
\paragraph{A model clumped in density.} First we consider a two component
density structure consisting of clumps and a rarefied inter-clump
medium (hereafter ICM), but keep the $\beta=1$ velocity law.  Clumps
are released randomly in radial direction at the inner boundary,
independently from each slice. The release in radial direction means
that a given clump stays within the same slice during its propagation
through the wind. The average time interval between the release of two
clumps is $\delta t$, which here and in the following is expressed in
units of the wind's dynamic time scale $t_{\rm dyn}=R_{\star}/\vinf$.

The average distance between clumps thus is $v_{\beta}\, \delta t$,
i.e.  clumps are spatially closer in the inner wind than in the outer
wind, and for example $\delta t = 0.5$ (in $t_{\rm dyn}$) gives an
average clump separation of 0.5 (in $R_{\star}$) at the point where $v
= 1$ (in $\vinf$).  We further assume that the clumps preserve mass
and lateral angle when propagating outwards, and that the underlying
model's total wind mass is conserved within every slice.  This radial
clump \textit{distribution} is the same as the one used by
\citet{Oskinova06} when simulating X-ray emission from O-stars, but
differs from the one used by \citet{Oskinova07} when investigating
porosity effects on resonance lines (see discussion in
Sect.~\ref{oskow}).  The radial clump \textit{widths} are here
calculated from the actual wind geometry and clump distribution by
assuming a \textit{volume filling factor} $f_{\rm v}$, defined as the
fractional volume of the dense gas\footnote{We here notice that
  $f_{\rm v}$ is normalized to the \textit{total} volume, i.e.,
  $f_{\rm v} = 0 \dots 1$. In some literature $f_{\rm v}$ is
  identified with the straight volume ratio $V_{\rm cl}/V_{\rm ic}$,
  which then implicitly assumes that $V_{\rm cl} \ll V_{\rm ic}$.}.  A
related quantity is the \textit{clumping factor}
\begin{equation}
	f_{\rm cl} \equiv \frac{\langle \rho^2 \rangle}{\langle \rho \rangle^2},
	\label{Eq:fcl}
\end{equation}
as defined by \citet{Owocki88}, where angle brackets denote temporal
averages.  Identifying temporal with spatial averages one may write for a
two component medium \citep[cf.][]{Abbott81} 
\begin{equation} 
f_{\rm cl} = \frac{f_{\rm v}+(1-f_{\rm v})\fic^2}{[f_{\rm v}+(1-f_{\rm v})\fic]^2},
\label{Eq:fclfv}
\end{equation} 
with
\begin{equation}
\fic \equiv \frac{\rho_{\rm ic}}{\rho_{\rm cl}},
\label{Eq:fic}
\end{equation} 
the ratio of low- to high-density gas (subscript ic denotes inter-clump and
cl denotes clump). For a void ($\fic=0$) ICM, $\rho_{\rm cl}/ \langle \rho
\rangle$ = $f_{\rm v}^{-1}=f_{\rm cl}$, i.e, $f_{\rm cl}$ then describes the
over density of the clumps as compared to the mean density. 

\begin{figure}
  \includegraphics[angle=90,width=6cm]{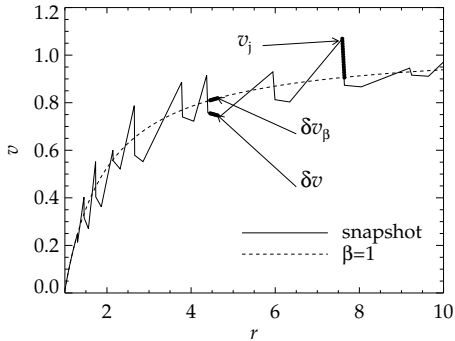}    
  \centering
  \caption{Non-monotonic velocity field and corresponding parameters in a 
    stochastic model.} 

\label{Fig:fclvcl}
\end{figure}

\paragraph{A model clumped in density and velocity.} Next we consider also a
non-monotonic velocity law, using the spatial distribution and widths of the
clumps described in the previous paragraph. The RH simulations indicate
that, generally, strong shocks separate denser and slower material from
rarefied regions with higher velocities. Building on this basic result, we
now modify the velocity fields in our stochastic models by adding a random
perturbation to the local $v_{\beta}$ value prior to the starting point of
each clump, so that the new velocity becomes $v_{\rm pre}$.  A `jump
velocity' is thereafter determined by a random subtraction from $v_{\beta}$,
now using the added perturbation as the maximum subtraction. That is, 
\begin{equation}
  v_{\rm pre}  = v_{\beta}+v_{\rm j} \times 2R_{\rm 1} \qquad
  v_{\rm post} = v_{\beta}-v_{\rm j} \times 2R_{\rm 1}R_{\rm 2},
  \label{Eq:vj}
\end{equation}
where $R_{\rm 1}$ and $R_{\rm 2}$ are two random numbers in the
interval 0 to 1. $v_{\rm pre}-v_{\rm post}$ is the jump velocity as
determined by the parameter $v_{\rm j}$.  By multiplying $R_{\rm 1}$
by two, we make sure that the mean perturbation at the `pre' point is
$v_{\rm j}$, and $R_{\rm 2}$ allows for an asymmetry about $v_{\beta}$
(see Fig.~\ref{Fig:fclvcl}). The clump is assumed to start at $v_{\rm
  post}$, and its velocity span is set by assuming a value for $\delta
v/\delta v_{\beta}$, where $\delta v$ is the velocity span of the
clump and $\delta v_{\beta}$ the corresponding quantity for the same
clump with a smooth velocity law (see
Fig.~\ref{Fig:fclvcl}). Inspection of our RH models suggests that
velocity gradients within density enhancements primarily are negative
(see also Sect.~\ref{vgrad}), and negative gradients are also adopted
in most of our stochastic models.  Finally we assume a constant
velocity gradient through the ICM.

Overall, the above treatment provides a phenomenological description
of the non-monotonic velocity field seen in RH simulations.  The
description differs from the one suggested by \citet{Owocki08}, who
uses only one parameter to characterize the velocity field (whereas we
have two). Our new formulation is motivated by both observational and
modeling constraints from strong and intermediate lines, as discussed
in Sect.~\ref{oskow}.

The basic parameters defining a stochastic model are listed in
Table~\ref{Tab:par}.  Fig.~\ref{Fig:contours} (right panel) shows the
density and velocity structures of one slice in a stochastic model,
with density parameters $f_{\rm v}=0.1$, $\delta t = 1.0$,
$\fic=0.005$, and velocity parameters $v_{\rm j}=0.15 v_\beta$ and
$\delta v = -\delta v_{\beta}$.  Clump positions have been highlighted
with filled dots and a comparison to a RH model (FPP) is given. In the
RH model, we have identified clump positions by highlighting all
density points with values higher than the corresponding smooth
model. The left panel shows the density contours of the same models,
where, for clarity, only the wind to $r=5$ is displayed.

\begin{table}
	\centering
	\caption{Basic parameters defining a stochastic wind model clumped in density and with a
          non-monotonic velocity field.}
	\tabcolsep=1.7mm
	\begin{tabular}{p{3.7cm}lp{0.6cm}l}
	  \hline \hline Name & Parameter &
          \multicolumn{2}{l}{Considered range} \\ \hline 
	  Volume filling factor & $f_{\rm v}$ & $f_{\rm v}$ & $= 0.01
          \dots 1.0$ \\ 
	  Average time interval~between release of~clumps & $ \delta t$ & 
	  $\delta t\,[t_{\rm dyn}]$ & $= 0.05 \dots 1.5$ \\ 
	  ICM density parameter, Eq.~\ref{Eq:fic} & $\fic$ & $\fic$ & $= 0 \dots 0.1$\\ 
	  Velocity~span~of~clump & $\delta v$ & $\delta v /
          \delta v_{\beta}$ & $= -10.0 \dots 1.0$ \\ 
	  Parameter determining the jump velocity & $v_{\rm j}$ & 
	  $v_{\rm j} / v_{\beta}$ & $= 0.01 \dots 0.15$ \\ 
	  \hline
	\end{tabular}
	\label{Tab:par}
\end{table}

\section{Radiative transfer} 
\label{rt} 
To compute synthetic line profiles from the wind models, we have
developed a Monte-Carlo radiative transfer code (MC-2D) that treats
resonance line formation in a spherical and axially symmetric wind
using an `exact' formulation (e.g., without resorting to the Sobolev
approximation).  The restriction to 2D is of course a shortage, but
has certain geometrical and computational advantages and should be
sufficient for the study of general properties, as discussed in
Sect.~\ref{3d}.  A thorough description and verification of the code
can be found in Appendix~\ref{rt_code}.

Photons are released from the lower boundary (the photosphere) and
each path is followed until the photon has either left the wind or
been backscattered into the photosphere.  Basic assumptions are a
line-free continuum with no limb darkening emitted at the lower
boundary, no continuum absorption in the wind, pure scattering lines,
instantaneous re-emission, and no overlapping lines (i.e., singlets).
These simplifying assumptions, except for doublet formation, are all
believed to be of minor importance to the basic problem.  By the
restriction to singlet line formation we avoid confusion between
effects on the line profiles caused by line overlaps and by other
important parameters, but on the other hand it also prevents a direct
comparison to observations for many cases (but see
Sect.~\ref{cmp_obs}). A consistent treatment of doublet formation will
be included in the follow-up study.

\section{First results from 2D inhomogeneous winds}
\label{2d}

\begin{figure}
  \begin{minipage}{8.8cm}
    \resizebox{\hsize}{!}
              {\includegraphics[angle=90]{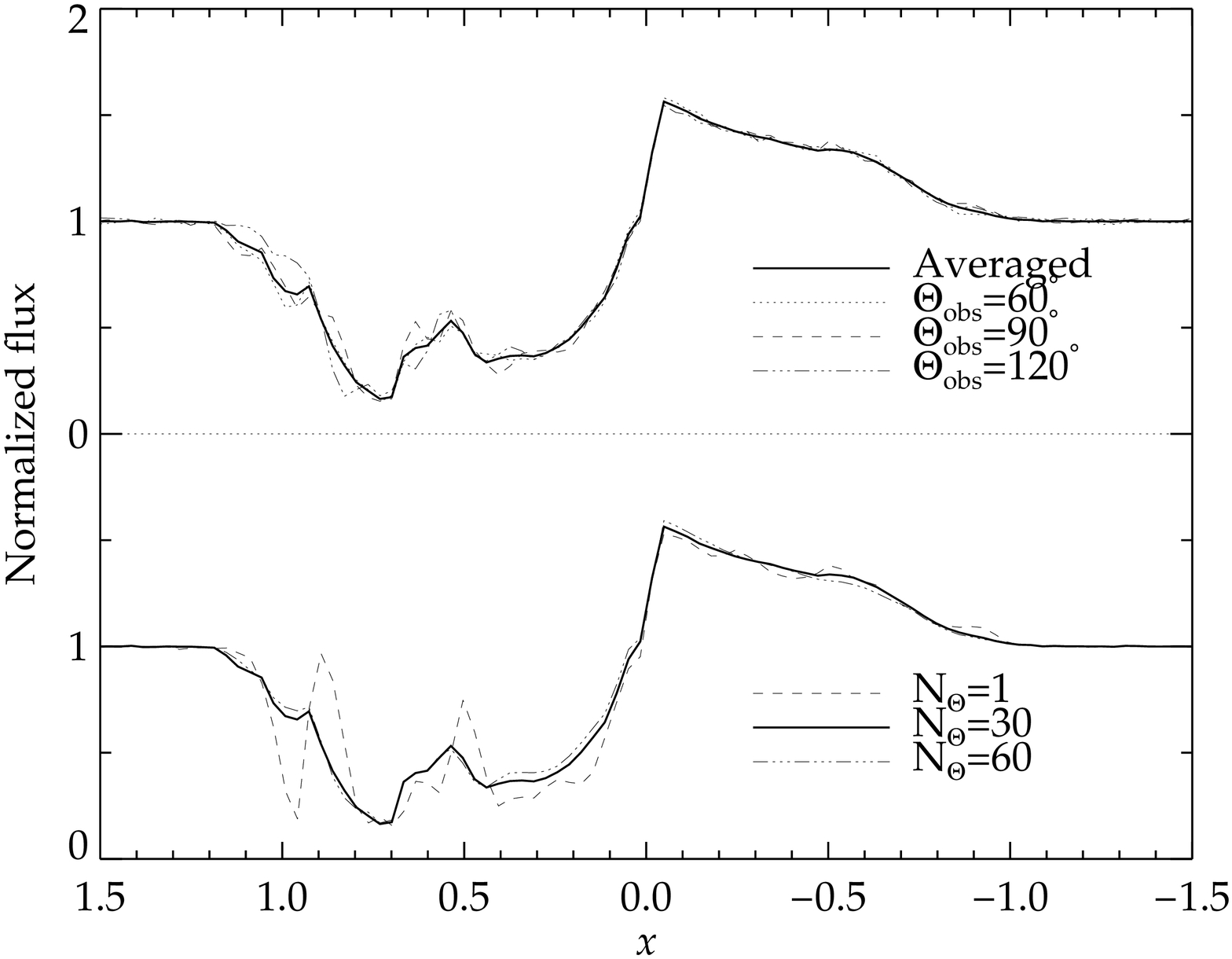}}
  \end{minipage}
  \begin{minipage}{8.8cm}
    \resizebox{\hsize}{!}
              {\includegraphics[angle=90]{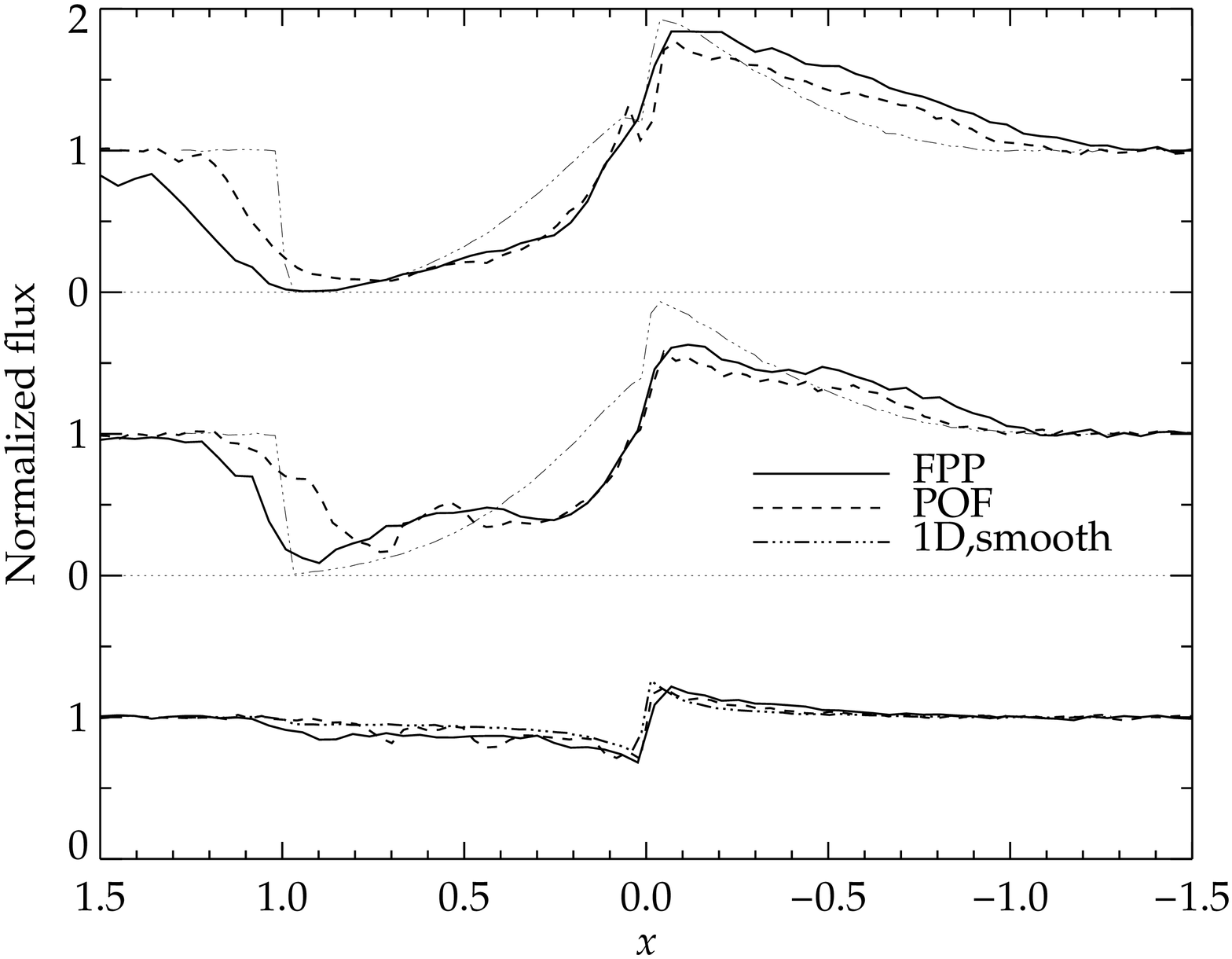}}
  \end{minipage}
  \caption{Synthetic line profiles calculated from 2D RH models.  The
    abscissa is the dimensionless frequency $x$ (Eq.~\ref{Eq:x}),
    normalized to the terminal velocity, and the ordinate is the flux
    normalized to the continuum. \textit{Upper panel:} Profiles from
    POF models with $\kappa_{\rm 0}=5.0$. The upper plot displays
    profiles for an observer placed at the $\Theta_{\rm obs}$ angles
    as labeled in the figure and a profile averaged over all
    $N_{\Theta}=30$ angles.  The lower plot displays averaged profiles
    for three different $N_{\Theta}$. \textit{Lower panel:} Averaged
    profiles from FPP and POF models with $N_{\Theta}=30$, and with
    $\kappa_{\rm 0}=100.0$ (upper) $\kappa_{\rm 0}=5.0$ (middle), and
    $\kappa_{\rm 0}=0.05$ (lower). For comparison, 1D, smooth profiles
    with the same values of $\kappa_0$ are shown as well.}
    \label{Fig:profs}
\end{figure} 

Throughout this section we assume a thermal velocity, $v_{\rm t} = 0.005$
(in units of $\vinf$ and $\sim 10 \ \rm km\,s^{-1}$, appropriate for a
standard O-star wind), and apply no microturbulence. After a brief
discussion on the impact of the observer's position and opening angles, we
concentrate on investigating the 
formation of strong, intermediate, and weak lines. In our
definition, an intermediate line is characterized by a line
strength\footnote{with $\kappa_0$ proportional to the product of mass-loss
rate and abundance of the considered ion, see Appendix~\ref{rt_code}.}
$\kappa_0 = 5.0$ chosen 
such as to almost precisely reach the saturation limit in a {\it smooth}
model (cf. Fig.~\ref{Fig:profs}).

By investigating these different line types, we account for the tight
constraints that exist for each flavor: i) \textit{weak lines} should
be independent of density-clumping properties as long as the clumps
remain optically thin, ii) for \textit{intermediate lines} either
smooth models overestimate the profile strengths or mass-loss rates
are lower than previously thought (e.g. the PV problem, see
Sect.~\ref{Introduction}), and iii) \textit{strong saturated lines}
are clearly present in hot star UV spectra, and observed features need
to be reproduced, such as high velocity ($> \vinf$) absorption, the
black absorption trough, and the reduction of re-emitted flux blueward
of the line center.

\subsection{Observer's position and opening angles}
\label{ang_dep} 
The observed spectrum as calculated from a 2D wind structure depends
on the observer's placement relative to the star (see
Appendix~\ref{rt_code}). As it turns out, however, this dependence is
relatively weak in both the stochastic and the RH models (the latter
is demonstrated in the upper panel of Fig.~\ref{Fig:profs}). Tests
have shown that the variability of the line profile's emission part is
insignificant.  The variability of the absorption part may be
detectable, at least near the blue edge, but is still insignificant
for the integrated profile strength; the equivalent width of the
absorption part is almost independent of the observer's position. Also
the opening angle, $180^{\circ}/N_{\Theta}$, primarily has a smoothing
effect on the profiles.  In Fig.~\ref{Fig:profs}, prominent discrete
absorption features appear near the blue edge in the model with
$N_{\Theta}=1$ (spherical symmetry), but are smoothed out in the
`broken-shell' models with $N_{\Theta}=30$ and $60$. The equivalent
widths of the absorption parts are approximately equal for all three
models.

Because our main interest here is the general behavior of the line
profiles, we choose to work only with $N_{\Theta}=30$ and profiles
averaged over all observer angles from here on.  Working with averaged
line profiles has great computational advantages, because roughly a
factor of $N_{\Theta}$ fewer photons are needed.

\subsection{Radiation-hydrodynamic models}
\label{rh}

Fig.~\ref{Fig:profs} (lower panel) shows line profiles from FPP and
POF hydrodynamical models. For the strong lines, the constraints
stated in the beginning of this section are reproduced without
adopting a highly supersonic and artificial microturbulence.  These
features arise because of the multiple resonance zones in a
non-monotonic velocity field, and are present in spherically symmetric
RH profiles as well (see POF for a comprehensive discussion); the main
difference between 1D and 2D is a smoothing effect, partly stemming
from averaging over all observer angles (see above).  The absorption
at velocities higher than the terminal is stronger in FPP than in POF,
due to both a higher velocity dispersion and a larger extent of the
wind ($r_{\rm max} \sim 30$ as compared to $r_{\rm max} \sim 5$, see
Sect.~\ref{wind_rh}); more overdense regions are encountered in the
outermost wind, which (because of the flatness of the velocity field)
leads to an increased probability to absorb at almost the same
velocities.

For the intermediate lines, we again see the qualitative features of
the strong lines, though less prominent. As compared to smooth models,
a minor \textit{absorption} reduction is present at velocities lower
than the terminal, but compensated by the blue edge
smoothing. Therefore the equivalent width of the line profile's
absorption part in the FPP model is approximately equal to that of the
smooth model, whereas in the POF model it is reduced by $\sim 10
\%$. This minor reduction agrees with that found by \citet{Owocki08},
and is not strong enough to explain the observations without having to
invoke a very low mass-loss rate.

For the weak lines, the absorption part is marginally stronger than from a
smooth, 1D model. 

\subsection{Stochastic models} 
\label{st}      

\begin{table*}
  \centering
  \caption{Primary stochastic wind models and parameters}
  \begin{tabular}{llllllll}
    \hline \hline
    Model name & $f_{\rm v}$ & $\delta t \ [\rm t_{\rm dyn}] $
    & $\fic$ & $\delta v/\delta v_{\beta}$ 
    & $v_{\rm j}/v_{\beta}$ & 
    $r_{\rm st}$$^{\rm a}$ & $r_{\rm ext}$$^{\rm b}$ \\ 
    \hline
    Default & 0.25 & 0.5 & 0.0025 & -1.0 & 0.15 & 1.3 & $\sim$\,25 \\
    RHcopy & 0.1 & 0.5 & 0.005 & -10.0 & 0.15 & 1.3 & $\sim$\,5 \\
    Obs1 & 0.11 & 0.5,4.0\,$^{\rm c}$ & 0.005,0.0025$^{\rm c}$ & -1.0 & 0.15 & 1.02 & $\sim$\,25 \\
    \hline
    \multicolumn{8}{l}{}                                   \\        
    \multicolumn{8}{l}{$^{\rm a}$ Radial onset of clumping. $^{\rm b}$ Radial extent of wind.} \\
\multicolumn{8}{l}{$^{\rm c}$ Left value inside the radius
corresponding to $v_{\beta}=0.6$, right value outside.}        \\
  \end{tabular}
  \label{Tab:mod}
\end{table*}
\begin{figure*}
  \sidecaption
     {\includegraphics[width=10.9cm,angle=90]{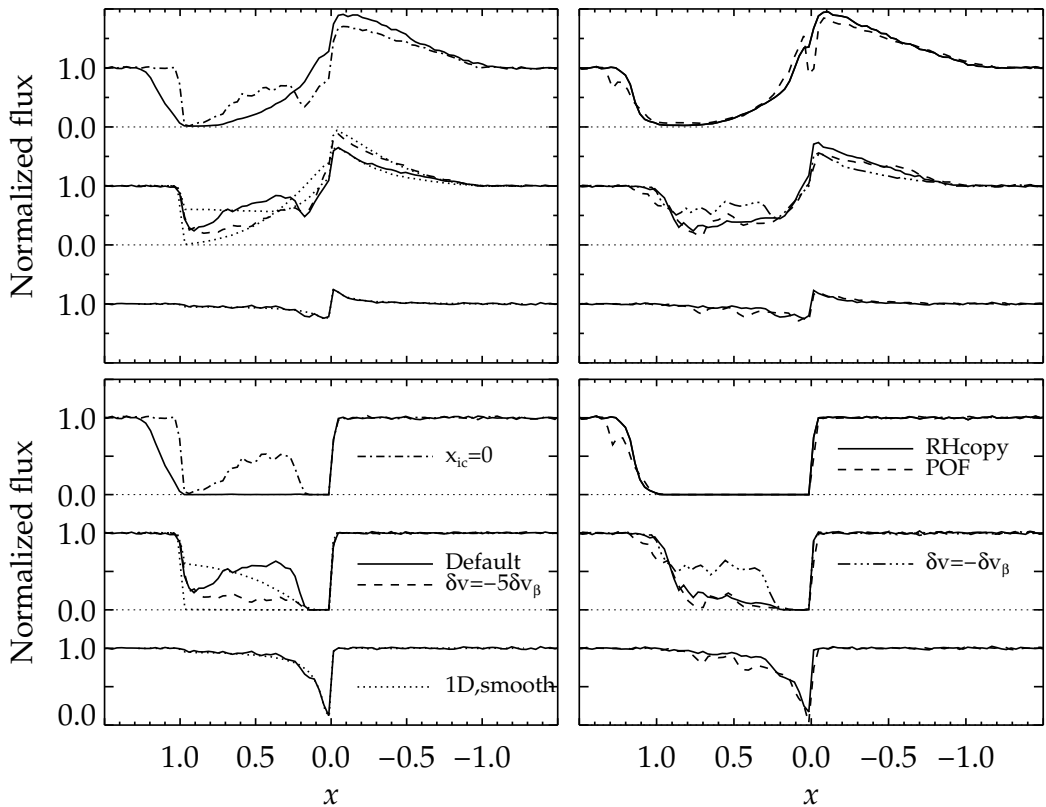}}        
      \caption{\textit{Left panels:} Solid lines display total line profiles
	and the absorption part for the default stochastic model (see
	Table~\ref{Tab:mod}), with $\kappa_0=1000$ (upper), $\kappa_0=5.0$
	(middle), and $\kappa_0=0.05$ (lower).  Dotted lines display smooth
	models with $\kappa_0=5.0$ and $\kappa_0=0.5$ (middle), and
	$\kappa_0=0.05$ (lower). Dashed/dashed-dotted lines with
	modifications from the default model as labeled in the figure.
	\textit{Right panels:} Same as the left panels, but for POF (dashed
	lines) and RHcopy (solid lines). Dashed-dotted lines with
	modifications from RHcopy as labeled in the figure.}
	\label{Fig:stochs}
\end{figure*} 

In this subsection we use a `default' 2D, stochastic model with parameters
as specified in Table~\ref{Tab:mod}. By comparing this model to models in
which one or more parameters are changed, we demonstrate key effects in the
behavior of the line profiles.

\paragraph{Strong lines.} For strong lines, the line profiles from the
default model reproduce the observational constraints described in the first
paragraph of this section.  As in the RH models, we apply no
microturbulence.  Fig.~\ref{Fig:stochs} (left panels) demonstrates the
importance of the ICM in the default model; the absorption part of a very
strong line is not saturated when $\fic=0$. That is, with a void ICM we
will, regardless of the opacity, always have line photons escaping their 
resonance zones without ever interacting with any matter, thereby
de-saturating the line.  This ICM finding agrees with that of
\citet{Zsargo08}, who point out that a non-void ICM is crucial for the
formation of highly ionized species such as O\,VI. We also notice that
$\delta v = - \delta v_{\beta}$ (used in the default model) does not permit
clumps to have velocities higher than the local $v_\beta$ value, preventing
absorption at velocities higher than the terminal one when the ICM is void.

\paragraph{Intermediate lines.} For intermediate lines, the line profiles
from the default model display the main observational requirement if to
avoid a drastic reduction in `smooth' mass-loss rates\footnote{Recall that
$f_{\rm v}=0.25 \rightarrow f_{\rm cl} \approx 4$, which implies
$\dot{M}=\dot{M}_{\rm smooth}/2$, if $f_{\rm cl}$ were derived from
$\rho^2$-diagnostics assuming optically thin clumps.}, namely a strong
absorption reduction as compared to a smooth model.  The left panels
of Fig.~\ref{Fig:stochs} show how the integrated profile strength of the
default model with $\kappa_0=5.0$ roughly corresponds to that of a smooth
model having $\kappa_0=0.5$, i.e., the smooth model would result in a
mass-loss rate (as estimated from the integrated profile strength) ten times
\textit{lower} than the clumped model. The figure also illustrates how the
main effect is on the absorption part of the line profile. In addition to
the reduction in profile \textit{strength}, the profile \textit{shapes} of 
the absorption parts are noticeably different for the default and smooth
models (the shapes of the re-emission parts, not shown here, are similar 
for the two models). We further discuss the shapes of the profiles in
Sect.~\ref{shapes}. The dramatic reduction in integrated profile strength
occurs because of large velocity gaps between the clumps, in which the wind
is unable to absorb (at this opacity the ICM may not `fill in' these gaps
with absorbing material). 

We have identified $|\delta v|$ as a critical parameter for the formation 
of intermediate lines. The importance of the velocity spans of the clumps 
is well illustrated by the absorption part profiles in Fig.~\ref{Fig:stochs}
(lower-left panel, middle plot). The absorption is much stronger in the
comparison model with $\delta v=-5 \delta v_{\beta}$ than in the default
model with $\delta v=-\delta v_{\beta}$, because the former model covers
more of the total velocity space \textit{within} the clumps, thereby closing
the gaps \textit{between} the clumps. Consequently the wind may, on average,
absorb at many more wavelengths. 

In principle, however, this effect is counteracted by a decrease in the
clump's optical depths, because of the now higher velocity gradients
($|\delta v/\delta v_{\beta}| >1$). Consider the \textit{radial} Sobolev
optical depth (proportional to $\rho/|\partial v/\partial r|$, see
Appendix~\ref{rt_code}) in a stochastic wind model. As compared to a smooth
model, the density inside a clump is enhanced by a factor of $f_{\rm
v}^{-1}$ (assuming a negligible ICM), but also the velocity gradient is
enhanced by a factor of $|\delta v/\delta v_\beta|$. Thus we may write for
the radial Sobolev optical depth inside a clump,
\begin{equation}
\tau_{\rm Sob} \approx \frac{\tau_{\rm Sob,sm}}{f_{\rm v}|\delta v/\delta v_\beta|}
\approx \frac{\kappa_{\rm 0}}{v_{\beta}f_{\rm v}|\delta v/\delta v_\beta|},
\label{Eq:tau_s}
\end{equation}
where `sm' indicates a quantity from a smooth wind, and the
expression to the right is valid for an underlying $\beta~=~1$ velocity
law. From Eq.~\ref{Eq:tau_s}, we see how the effects on the optical depth
from the increased density ($f_{\rm v }=0.25$) and the increased velocity
gradients ($|\delta v/\delta v_{\beta}|=5$) almost cancel each other in this
example.  Thus, the clumps are still optically thick for the intermediate
line ($\kappa_0=5$), which means that the larger coverage of the total
velocity space `wins', and the net effect becomes an increase in absorption
(as seen in Fig.~\ref{Fig:stochs}, lower-left panel, middle plot). This will
be true as long as not $f_{\rm v}|\delta v/\delta v_\beta| \gg 1$, 
which is never the case in the parameter range considered here.

Finally, the prominent absorption dip toward the blue edge in the default
model turns out to be a quite general feature of our stochastic models, and
is discussed in Sects.~\ref{eta} and \ref{outer}. 

\paragraph{Weak lines.} The statistical treatment of density clumping
included in atmospheric codes such as CMFGEN, PoWR, and FASTWIND is
valid for optically thin clumps and a negligible ICM, and gives no
direct effect on resonance lines scaling linearly with density. Here
we test this prediction using detailed radiative transfer\footnote{The
  \textit{indirect} effect through the feedback on the occupation
  numbers is not included, because in this section we assume constant
  ionization.}.  Our default model recovers the smooth results when
$\kappa_{\rm 0}=0.05$ (Fig.~\ref{Fig:stochs}, left panels), confirming
the expected behavior.  However, from calculating spectra using
different values of $\kappa_0$, we have found that significant
deviations from smooth models occur for the default model already
before $\kappa_{\rm 0}$ reaches unity. This occurs because the clumps
start to become optically thick, which may again be understood by
considering the radial Sobolev optical depth (Eq.~\ref{Eq:tau_s}).
With $f_{\rm v} \leq 0.25$ and $\kappa_0 \geq 0.25$, one finds
$\tau_{\rm Sob} \geq 1.0$.

\begin{figure}
  \resizebox{\hsize}{!}{\includegraphics[angle=90]{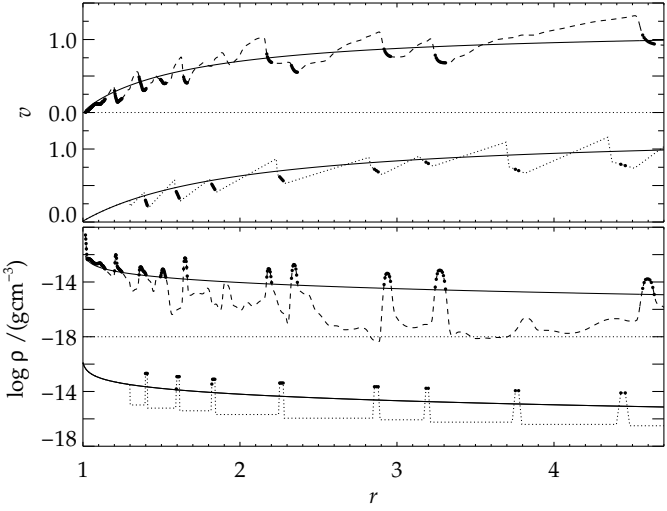}} 
    \caption{Velocity (\textit{upper panel}) and density
      (\textit{lower panel}) structures for one slice in POF (dashed)
      and RHcopy (dotted), see Table~\ref{Tab:mod}.  Solid lines are
      the corresponding smooth structures, and clumps are highlighted
      as in Fig.~\ref{Fig:contours}.}
    \label{Fig:hs}
\end{figure}

\subsection{Comparison between stochastic and radiation-hydrodynamic models} 

\label{cmp} Our stochastic wind models have been constructed to contain all
essential ingredients of the RH models. Therefore they should also 
reproduce the RH results, at least qualitatively, if a suitable parameter
set is chosen. To test this we used the POF model. In this model, the
clumping factor increases drastically at $r \sim 1.3$, from $f_{\rm
  cl} \sim 1.0$ to $f_{\rm cl} \sim 10$, after which it stays
basically constant. The average clump separation in the outer wind is
roughly half a stellar radius.  Important for the velocity field is
that the velocity spans of the clumps are generally \textit{larger}
than corresponding `$\beta$ spans', i.e., $|\delta v/\delta v_{\beta}|
> 1$ (this is the case in FPP as well), a characteristic behavior that
primarily affects the intermediate lines (details will be discussed in
Sect.~\ref{vgrad}).  Finally, a suitable $v_{\rm j}$ can be assigned
from the position of the blue edge in a strong line calculated from
POF. Table~\ref{Tab:mod} (entry RHcopy) summarizes all parameters used
to create this stochastic, `pseudo-RH' model. Fig.~\ref{Fig:hs}
displays one slice of the velocity and density structures in the POF
and RHcopy models, and Fig.~\ref{Fig:stochs} (right panels) displays
the line profiles.

The line profiles of POF are matched reasonably well by RHcopy.  The
intermediate lines again demonstrate the importance of the velocity spans of
the clumps; for an alternative model with $\delta v = -\delta v_{\beta}$,
there is much less absorption in the stochastic model than in POF, i.e., we
encounter the same effect as discussed in the previous subsection.  We
conclude that in RH models it is the large velocity spans inside the density
enhancements that prevent a reduction in profile strength (as compared to 
smooth models) for intermediate lines. 

\section{Parameter study}
\label{ps}           

Having established basic properties, we now use our stochastic models
to analyze the influence from different key parameters in more detail.
First, however, we introduce a quantity that turns out to be
particularly useful for our later discussion.

\subsection{The effective escape ratio}
\label{eta}

\begin{figure}
  \resizebox{\hsize}{!}{\includegraphics[angle=90]{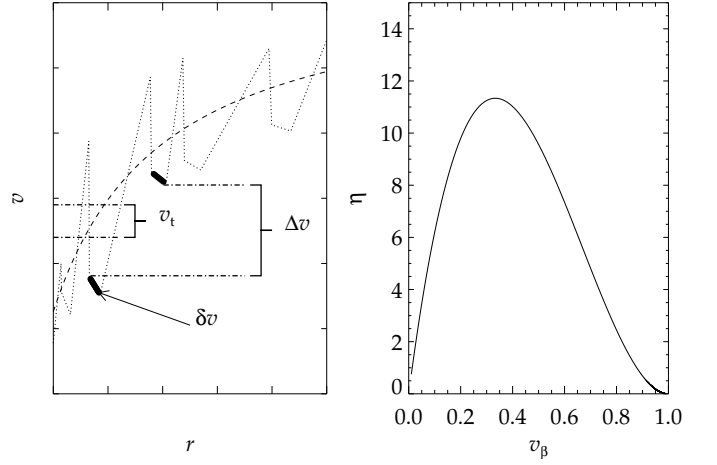}} 
    \caption{\textit{Left:} Schematic of $\Delta v$, the velocity gap
    between two subsequent clumps, made up by velocities not covered by
    \textit{any} of the clumps. $\delta v$ is the velocity span of a clump
    and $v_{\rm t}$ the thermal velocity. \textit{Right:} The effective 
    escape ratio $\eta$ (Eq.~\ref{Eq:heff}) as a function of $\beta=1$
    velocity, for the parameters of the default model (see
    Table~\ref{Tab:mod}).} 
    \label{Fig:fer} 
\end{figure}

For the important intermediate lines, it is reasonable to assume that
the clumps are optically thick and the ICM negligible (see
Sect.~\ref{st} and the next paragraph). Under these assumptions, a
decisive quantity for photon absorption will be the velocity gap {\it
  not} covered by the clumps, as compared to the thermal velocity (the
latter determining the width of the resonance zone in which the photon
may interact with the wind material). This is illustrated in the left
panel of Fig.~\ref{Fig:fer}, and we shall call this quantity the
`effective escape ratio'
\begin{equation} 
  \eta \equiv \frac{\Delta \it v}{v_{\rm t}},
\label{Eq:fer}
\end{equation}
where $\Delta v$ is the velocity gap between two subsequent clumps,
made up by all velocities not covered by \textit{any} of the clumps
(see Fig.~\ref{Fig:fer}). In principle, $\eta$ determines to which
extent the vorosity effect \citep[i.e., the velocity gaps between the
  clumps, cf.][]{Owocki08} is important for the line formation.  As
defined, $\eta$ does not contain any assumptions on the
\textit{spatial} structure of the wind.  $\eta << 1$ means that the
velocity gaps between the clumps are much smaller than the thermal
velocity, which in turn means that the probability for a photon to
encounter a clump within its resonance zone is high.  If we assume
each clump to be optically thick, every encounter will lead to an
absorption. Thus the probability for photon absorption is high when
the value of $\eta$ is low. Vice versa, $\eta >> 1$ results in a high
probability for the photon to escape its resonance zone without
interacting with the wind material, i.e., a low absorption
probability. If the entire velocity space were covered by clumps,
$\eta=0$.

For the wind geometry used in our stochastic models, we may write (see
Appendix~\ref{app_eta} for a derivation)
\begin{equation} 
  \eta \approx \frac{v_{\beta}\delta t(1-f_{\rm v}|\delta v/\delta v_{\beta}|)}{L_{\rm r}} \approx
  \frac{\delta t (1-f_{\rm v}|\delta v/\delta v_{\beta}|)}{v_{\rm t}}\frac{v_{\beta}}{r^2},
  \label{Eq:heff}
\end{equation}  
where $L_{\rm r}$ is the radial Sobolev length of a smooth model,
which for $\beta=1$ is $L_{\rm r} \approx v_{\rm t}r^2$ (as usual, $r$
and $L_{\rm r}$ in $R_\star$ and $\delta t$ in $t_{\rm dyn}$). Note
that in Eq.~\ref{Eq:heff} also the density-clumping parameters have
entered the expression for $\eta$, illustrating that there is an
intimate coupling with the \textit{spatial} clumping parameters, even
though the vorosity effect initially depends on velocity parameters
alone.  For example, consider a wind with clumps that follow a smooth
$\beta$ velocity law. By bringing the clumps spatially closer together
(for example by decreasing $\delta t$), the velocity gaps between them
decrease as well. Thus one may choose to describe the changed
situation either in terms of a less efficient porosity, because of
fewer `density holes' in the resonance zone through which the photons
can escape \citep[as done by][]{Oskinova07}, \textit{or} in terms of a
less efficient vorosity, because of smaller velocity gaps between the
clumps.  Of course, one may also obtain a lower velocity gap between
the clumps by increasing the actual velocity spans inside the clumps,
as simulated in our stochastic models when $|\delta v/\delta
v_{\beta}| >1$.  This effect, leading to a rather low vorosity, has
already been demonstrated to be at work in the RH models
(Sect.~\ref{cmp}).
 
Using the parameters of our default model, Fig.~\ref{Fig:fer} (right panel) 
displays $\eta$ as a function of velocity and shows that $\eta$ increases
rapidly in the inner wind, reaches a maximum at $v \approx 0.33$, and then
drops in the outer wind.  To compare this behavior with that of the line
profiles, we can associate absorption at some frequency $x_{\rm obs}$ with
the corresponding value of the velocity, because absorption occurs at
$x_{\rm obs} \approx \mu v \approx v$ (radial photons dominate). In the
default model's absorption-part line profile (see Fig.~\ref{Fig:stochs}, 
the middle plot in the lower-left panel), a strong de-saturation occurs
directly after the clumping is set to start (at $r=1.3$, $v \approx 0.23$),
followed by a maximum at $x_{\rm obs} \approx 0.35$, and finally an
absorption dip toward the blue edge. The behavior of the line profile is
thus well mapped by $\eta$, and we may explain the absorption dip as a
consequence of the low value of $\eta$ in the outer wind, which in turn
stems from the slow variation of the velocity field (i.e., from radially
extended resonance zones). 

\subsection{Density parameters}
\label{dens_par}

\begin{figure}
  \centering
 \resizebox{\hsize}{!}{\includegraphics[angle=90]{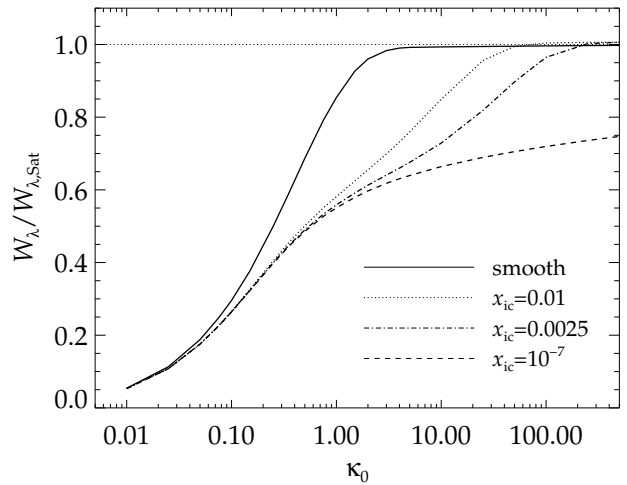}} 
      \caption{Equivalent widths $W_{\lambda}$ of the absorption
      parts of line profiles, normalized to the value of a saturated line,
      as a function of line strength parameter $\kappa_{\rm 0}$. The solid
      line is calculated from smooth models, and the dashed, dashed-dotted,
      and dotted lines from stochastic models with a smooth velocity field
      and $\delta t=0.5$, $f_{\rm v}=0.25$, and $\fic$ as indicated.}
      \label{Fig:ew_1d}
\end{figure}

To isolate density-clumping effects, we use a smooth $\beta =1$
velocity law in this subsection.  Despite the smooth velocity field,
there are still holes in velocity space (because of the density
clumping, at the locations where the ICM is present), and the
expression for $\eta$ (Eq.~\ref{Eq:heff}) remains valid.  Since a
smooth velocity field corresponds to $\delta v=\delta v_{\beta}$, also
the run of $\eta$ is equal to the one displayed in
Fig.~\ref{Fig:fer}. In this subsection we work only with integrated
profile strengths (characterized by the equivalent width $W_{\lambda}$
of the line's absorption part).  The shapes of the line profiles are
discussed in Sect.~\ref{shapes}.

Fig.~\ref{Fig:ew_1d} shows $W_{\lambda}$ as a function of $\kappa_0$,
for smooth models as well as for stochastic models with and without a
contributing ICM. The figure directly tells: i) The default model
($\fic = 0.0025$) for the intermediate line ($\kappa_0=5.0$) displays
a $W_{\lambda}$ corresponding to a smooth model with a $\kappa_0$
roughly ten times lower.  ii) Lines never saturate if the ICM is
(almost) void.  iii) The run of $W_{\lambda}$ for the smooth and
clumped models decouple well before $\kappa_0$ reaches unity.  iv) For
intermediate lines, the response of $W_{\lambda}$ on variations of
$\kappa_0$ is weak for clumped models.  Points one to three confirm
our findings from Sect.~\ref{st}.

A variation of $\delta t$ in the stochastic models affects 
primarily the high $\kappa_0$ part ($\kappa_0 \ga 1.0$) of the curves in
Fig.~\ref{Fig:ew_1d}. For example, lowering $\delta t$ in the model with a void ICM results in an upward shift 
of the dashed curve and vice versa. To obtain saturation with a void ICM,
$\delta t \approx 0.05$ is required, which may be understood in terms of
Eq.~\ref{Eq:heff}. 
For $\delta t = 0.05$, the $\eta$-values corresponding to the default model
are decreased by a factor of ten, and $\eta$ reaches a maximum of only about
unity, with even lower values for the majority of the velocity space (cf.
Fig.~\ref{Fig:fer}, right panel). The velocity gaps between the clumps then
become closed, and the line saturates. In this situation, however, the
intermediate line becomes saturated as well, again demonstrating the
necessity of a {\it non-void} ICM to simultaneously saturate a strong line
and not saturate an intermediate line. Only a properly chosen $\fic$
parameter ensures that the velocity gaps between the clumps become filled by
low-density material able to absorb at strong line opacities, but
\textit{not} (or only marginally) at opacities corresponding to intermediate
lines. 
  
When varying $\fic$, the primary change occurs at the high $\kappa_0$
end of Fig.~\ref{Fig:ew_1d}. For higher (lower) values of $\fic$, this
part becomes shifted to the left (right), and the curve decouples
earlier (later) from the corresponding curve for the void ICM. A
higher ICM density obviously means that the ICM starts absorbing
photons at lower line strengths and vice versa. Thus, observed
saturated lines could potentially be used to derive the ICM density
(or at least to infer a lower limit), \textit{if} the mass-loss rate
(and abundance) is known from other diagnostics.

The behavior of the absorption with respect to the volume filling
factor is as expected from the expression for $\eta$; the higher
$f_{\rm v}$, the lower the value of $\eta$, and the stronger the
absorption.  This is because a higher $f_{\rm v}$ for a fixed $\delta
t$ implies that the clumps become more extended, whereas the distances
between clump centers remain unaffected. Consequently, a larger
fraction of the total wind velocity is covered by the clumps, leading
to stronger absorption.  For weak lines ($\kappa_0 \approx 0.05$), the
ratio $W_{\lambda}/W_{\rm \lambda,sm}$ deviates significantly from
unity only when $f_{\rm v} \la 0.1$. Only for such low values can high
enough clump densities be produced so that the clumps start to become
optically thick.

\smallskip 
From Fig.~\ref{Fig:ew_1d} it is obvious that, generally, clumped
models have a different (slower) response in $W_{\lambda}$ to an
increase in $\kappa_0$ than do smooth models. This behavior may be
observationally tested using UV resonance doublets \citep{Massa08},
because the only parameter that differs between the two line
components is the oscillator strength. Thus, if a smooth wind model is
used and the fitted ratio of line strengths (i.e., $\kappa_{\rm
  0,blue}/\kappa_{\rm 0,red}$) does not correspond to the expected
ratio of oscillator strengths, one may interpret this as a signature
of a clumped wind. Such behavior was found by \citet{Massa08}, where
the observed ratios of the blue to red component of Si\,IV
$\lambda\lambda$1394,1403 in B supergiants showed a wide spread
between unity and the expected factor of two. This result indicates
precisely the slow response to an increase in $\kappa_0$ that is
consistent with inhomogeneous wind models such as those presented
here, but not with smooth ones. In inhomogeneous models, the expected
profile strength (or $W_{\lambda}$) ratio between two doublet
components will depend on the adopted clumping parameters (as
demonstrated by Fig.~\ref{Fig:ew_1d} and the discussion above) and may
in principle take any value in the range found by
\citeauthor{Massa08}. That is, while a profile-strength ratio
deviating from the value expected by smooth models might be a clear
indication of a clumped wind, the opposite is not necessarily an
indication of a smooth wind.  Furthermore, the degeneracy between a
variation of clumping parameters and $\kappa_0$ suggests that
un-saturated resonance lines should be used primarily as consistency
tests for mass-loss rates derived from other diagnostics rather than
as direct mass-loss estimators. We will return to this problem in
Sect.~\ref{cmp_obs}, where a first comparison to observations is
performed for the PV doublet.

\subsection{Velocity parameters}
\label{vel_par} 
\begin{figure}
  \resizebox{\hsize}{!}{\includegraphics[angle=90]{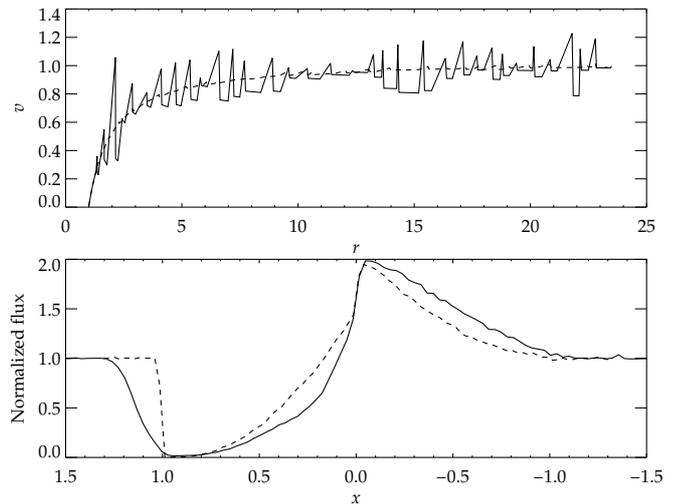}}              
  \caption{\textit{Upper:} Velocity structures (one slice) in 
    two stochastic models with density-clumping parameters as for the 
    default model, and different velocity parameters. Dashed:
    $\delta v/\delta v_\beta=1$ and $v_{\rm j}/v_\beta=0.01$. Solid: 
    $\delta v/\delta v_\beta=-1$ and $v_{\rm j}/v_\beta=0.5$ below
    $v_{\beta}=0.6$ and 
    $v_{\rm j}/v_\beta=0.15$ above. \textit{Lower:} Corresponding line
    profiles for a strong line.}
  \label{Fig:vj}
\end{figure}

The jump velocity parameter, $v_{\rm j}$, affects only the strong
lines (or, more specifically, the lines for which the ICM is
significant), and determines the maximum velocity at which absorption
can occur. For example, by setting $v_{\rm j}=0$, no absorption at
frequencies higher than $x=1$ is possible (unless $\delta v$ is
positive and very high).  A higher $v_{\rm j}$ also implies more
velocity overlaps, and thereby an increased amount of backscattering
due to multiple resonance zones. Both effects are illustrated in
Fig.~\ref{Fig:vj}.  Judging from the line profiles of the lower panel,
the blue edge and the reduction of the re-emitted flux blueward of the
line center may both be used to constrain $v_{\rm j}$.  The upper
panel shows one slice of the corresponding velocity fields,
illustrating that the underlying $\beta$ law is recovered almost
perfectly when using $v_{\rm j}=0.01v_\beta$ and $\delta v=\delta
v_\beta$. With this velocity law and a non-void ICM, the corresponding
strong line profile is equivalent to a profile from a smooth model.

In Sects.~\ref{st} and~\ref{cmp}, we showed that a higher value of the
clumps' velocity spans led to stronger absorption for intermediate
lines.  In principle this is as expected from Eq.~\ref{Eq:heff}, where
$\eta$ always decreases with increasing $|\delta v/\delta
v_\beta|$. However, with the very high value of $|\delta v/\delta
v_\beta|$ used in, e.g., the RHcopy model, one realizes that $\eta$ in
Eq.~\ref{Eq:heff} becomes identically zero, because $f_{\rm v}|\delta
v/\delta v_\beta|=1$.  An $\eta=0$ corresponds to the whole velocity
space being covered by clumps, and the saturation limit should be
reached. As is clear from Fig.~\ref{Fig:stochs}, however, this is not
the case. This points out two important details not included when
deriving the expression for $\eta$ and interpreting the absorption in
terms of this quantity, namely that clumps are distributed randomly
(with $\delta t$ determining only the average distances between them)
and that the parameter $v_{\rm j}$ allows for an asymmetry in the
velocities of the clumps' starting points (see
Sect.~\ref{wind_stoch}).  These two issues lead to overlapping
velocity spans for some of the clumps, whereas for others there is
still a velocity gap left between them, through which the radiation
can escape.  Therefore the profiles do not reach complete saturation,
despite that on average $\eta=0$.
This illustrates some inherent limitations when trying to interpret
line formation in terms of a simplified quantity such as $\eta$.

The impact from the velocity spans of the clumps on the line profiles
also depends on the density-clumping parameters. To achieve
approximately the same level of absorption, a higher value of $\delta
v/\delta v_{\beta}$ was required in the RHcopy model ($f_{\rm v}=0.1$)
than in the default model ($f_{\rm v}=0.25$), see
Fig.~\ref{Fig:stochs}.  Since $\delta v_{\beta} \propto f_{\rm v}
\delta t$ (see Appendix~\ref{app_eta}), the actual velocity spans of
the clumps are different for different density-clumping parameters,
even if $\delta v / \delta v_{\beta}$ remains unchanged.

By changing the sign of $\delta v$ in the default model (that is, assuming a
positive velocity gradient inside the clumps), we have found that our
results qualitatively depend only on $|\delta v|$. 
Some details differ though. For example, a $\delta v > 0$ in our stochastic
models permits absorption at velocities higher than the terminal one also
within the clumps, whereas $\delta v < 0$ restricts the clump velocities to
below the local $v_{\beta}$ (see Fig.~\ref{Fig:fclvcl}).  In this matter
$v_{\rm j}$ plays a role as well, since $v_{\rm j}$ controls where, with
respect to the local $v_{\beta}$, the clumps begin.  For reasonable values
of $v_{\rm j}$, however, its influence is minor on lines where the ICM is
insignificant.  Finally, tests have confirmed that optically thin lines are
only marginally affected when varying $\delta v/\delta v_{\beta}$.

\section{Discussion}
\label{Discussion}

\subsection{The shapes of the intermediate lines}
\label{shapes}

\begin{figure}
  \resizebox{\hsize}{!}{\includegraphics[angle=90]{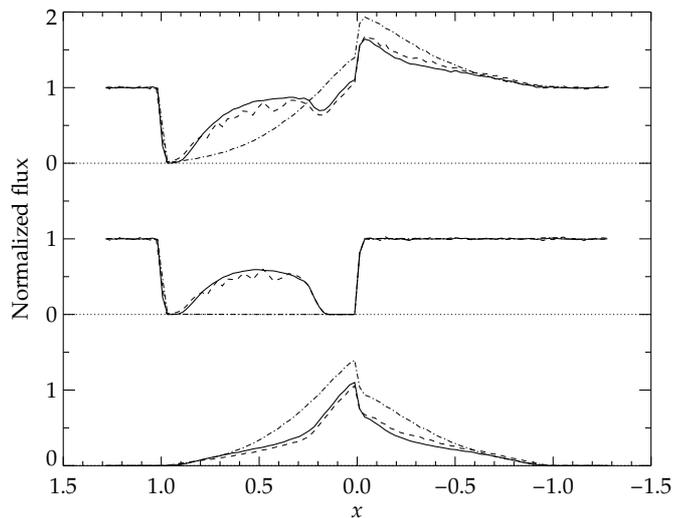}}   
  \caption{Total, absorption part, and re-emission part line profiles 
    for 1D, smooth models with $\kappa_0=5.0$ (dashed-dotted lines)
    and $\kappa_0=5.0/(2\eta)$ (solid lines, see Sect.~\ref{shapes}), 
    and for a 2D, stochastic model with density parameters as the 
    default model and a $\beta=1$ velocity law (dashed lines).}           
  \label{Fig:1d_fer}
\end{figure} 

For intermediate lines, the shape of the absorption part of the default
model differs significantly from the shape of a smooth model (see
Fig.~\ref{Fig:stochs}, the middle plot in the lower-left panel). We showed
in Sect.~\ref{eta} that the shapes could be qualitatively understood by the
behavior of $\eta$. 
This is further demonstrated here by scaling the line strength parameter of
a 1D, smooth model, using a parameterization $\kappa_0 \propto \eta^{-1}$
outside the radius $r=1.3$ where clumping is assumed to start.
Fig.~\ref{Fig:1d_fer} displays the line profiles of 1D, smooth models with
$\kappa_0=5.0$ and $\kappa_0=5.0/(2\eta)$. These profiles are compared to
those calculated from a `real' 2D stochastic model with density-clumping
parameters as the default model, but with a $\beta=1$ velocity field. $\eta$
was calculated from Eq.~\ref{Eq:heff}, using the parameters of the default
model and a $\beta=1$ velocity law, and the factor of 2 in the denominator
of the scaled $\kappa_0$ was chosen so that the \textit{integrated} profile
strength of the 2D model was roughly reproduced. From Fig.~\ref{Fig:1d_fer}
it is clear that the 1D model with scaled $\kappa_0$ well reproduces the 2D
results, indicating that indeed $\eta$ governs the shape of the line
profile. We notice also that these profiles display a completely black
absorption dip in the outermost wind, as opposed to the default model with a
non-monotonic velocity field (see Fig.~\ref{Fig:stochs}, the middle plot in
the lower-left panel).  This is because the $\beta$ velocity field does not
allow for any clumps to overlap in velocity space (see the discussion in
Sect.~\ref{vel_par}), making the mapping of $\eta$ almost perfect.

Let us also point out that the line shapes can be somewhat altered
by using a different velocity law, e.g., $\beta \ne 1$. Such a change would
affect the distances between clumps as well as the Sobolev length, and
thereby the line shapes of both absorption and re-emission profiles.
However, in all cases is the shape of the re-emission part similar in the
clumped and smooth models.

\subsection{The onset of clumping and the blue edge absorption dip}
\label{outer}

\begin{figure}
  \resizebox{\hsize}{!}{\includegraphics[angle=90]{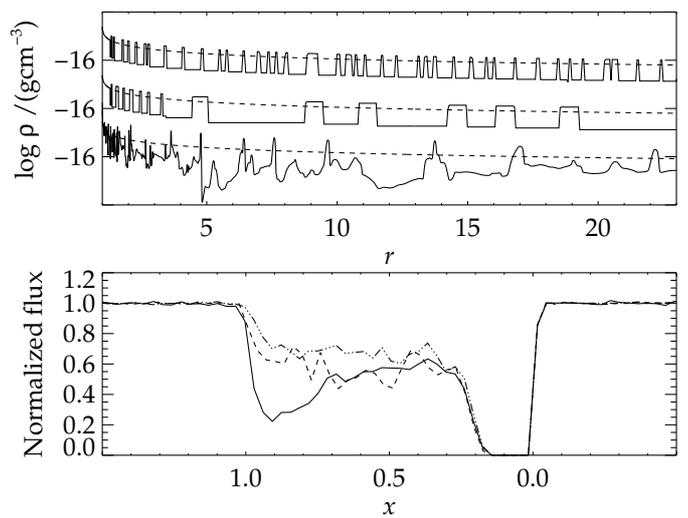}} 
    \caption{\textit{Upper panel:} Density structures of one slice in
      the default stochastic model (upper), in the default stochastic
      model with a modified $\delta t$ (middle, see
      Sect.~\ref{outer}), and in FPP (lower). \textit{Lower panel:}
      Line profiles for the absorption part of an intermediate line
      for the default model (solid line), for the default model with a
      modified $\delta t$ (dashed line), and for the default model
      with an ionization structure decreasing with increasing velocity
      (dashed-dotted line, see text).} \label{Fig:holes}
\end{figure} 

We have used $r=1.3$ as the onset of wind clumping in our stochastic
models, which roughly corresponds to the radius where significant
structure has developed from the line-driven instability in our RH
models.  However, \citet{Bouret03,Bouret05} analyzed O-stars in the
Galaxy and the SMC, assuming optically thin clumps, and found that
clumping starts deep in the wind, just above the sonic point. Also
\citet{Puls06} used the optically thin clumping approach, on
$\rho^2$-diagnostics, and found similar results, at least for O-stars
with dense winds.  With respect to our stochastic models, the
qualitative results from Sects.~\ref{2d}~and~\ref{ps} remain valid
when choosing an earlier onset of clumping. Quantitatively, the
integrated absorption in intermediate lines becomes somewhat weaker,
because the clumping now starts at lower velocities, and of course the
line shapes in this region are affected as well.  The onset of wind
clumping will be important when comparing to observations, as
discussed in Sect.~\ref{cmp_obs}.

The stochastic models that de-saturate an intermediate line generally
display an absorption dip toward the blue edge (see
Figs.~\ref{Fig:stochs} and \ref{Fig:1d_fer}), which has been
interpreted in terms of low values of $\eta$ in the outer wind (see
Sect.~\ref{eta}). However, this characteristic feature (not to be
confused with the so-called DACs, discrete absorption components) is
generally not observed, and one may ask whether it might be an
artifact of our modeling technique.  In the following we discuss two
possibilities that may cause our models to overestimate the absorption
in the outer wind; the ionization fraction and too low clump
separations.

Starting with the former, we have so far assumed a constant ionization
factor, $q=1$ (cf. Eq.~\ref{Eq:kappa0}). This is obviously an
over-simplification.  For example, an outwards decreasing $q$ would
result in less absorption toward the blue edge. Here we merely
demonstrate this general effect, parameterizing $q = v_0/v_{\beta}$ in
the stochastic default model (see Table~\ref{Tab:mod}), with $v_{\rm
  0}=0.1$ the starting point below which $q=1$.  Fig.~\ref{Fig:holes}
(lower panel, dashed-dotted lines) shows how the absorption in the
outer wind becomes significantly reduced.

The temperature structure of the wind is obviously important for the
ionization balance.  Whereas an isothermal wind is assumed in POF (see
Sect.~\ref{wind_rh}), the FPP model has shocked wind regions with
temperatures of several million Kelvin. To roughly map corresponding
effects on the line profiles, we re-calculated profiles based on FPP
models assuming $q=0$ in all regions with temperatures higher than
$T=10^5\rm\,K$, and $q=1$ elsewhere.  Since the hot gas resides
primarily in the low-density regions, however, the emergent profiles
were barely affected, and particularly intermediate lines remained
unchanged.

On the other hand, the X-ray emission from hot stars (believed to
originate in clump-clump collisions, see FPP) is known to be crucial
for the ionization balance of highly ionized species such as C\,IV,
N\,V, and O\,VI \citep[see, e.g., the discussion in][]{Puls08}. X-rays
have not been included here, but could in principle have an impact on
our line profiles, by illuminating the over-dense regions and thereby
changing the ionization balance. \citet{Krticka09}, however, find that
incorporating X-rays does not influence the PV ionization significantly.
Finally, non-LTE analyses including feedback from optically thin
clumping have shown that this as well has a significant effect on the
derived ionization fractions of, e.g., PV \citep[]{Bouret05,Puls08b}.
To summarize, it is clear that a full analysis of ionization fractions
must await a future non-LTE application that includes relevant
feedback effects from an inhomogeneous wind on the occupation numbers.

In RH models, the average distance between clumps increases in the
outer wind, due to clump-clump collisions and velocity stretching
\citep{Feldmeier97,Runacres02}.  Neglecting the former effect, our
stochastic models have clumps much more closely spaced in the outer
wind\footnote{The effect is minor in POF, since these RH models only
  extend to $r \sim 5$ (see Sect.~\ref{wind_rh}).}.  We have therefore
modified the default model by setting $\delta t = 3$ outside a radius
corresponding to $v_{\beta}=0.7$. This is illustrated in the upper
panel of Fig.~\ref{Fig:holes}. The mass loss in the new stochastic
model is preserved (because the clumps are more extended, see the
figure), and this model now better resembles FPP. Recall that
differences in the widths of the clumps are expected, since in the
default model $f_{\rm cl} \approx f_{\rm v}^{-1}=4$, whereas in FPP
$f_{\rm cl} \approx 10$. The corresponding line profile shows how the
absorption outside $x \approx 0.7$ has been reduced, as expected from
the higher $\delta t$.

\subsection{The velocity spans of the clumps}
\label{vgrad}

\begin{figure}
  \resizebox{\hsize}{!}{\includegraphics[angle=90]{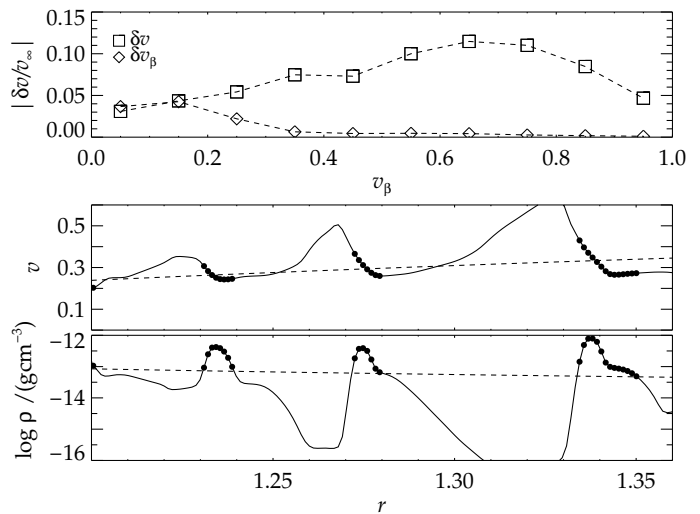}}              
  \caption{\textit{Upper:} Velocity spans of
    density enhancements in the FPP model (squares) and corresponding 
    $\beta$ intervals (diamonds). \textit{Lower:} Three density enhancements 
    and corresponding velocity spans in the FPP model, 
    highlighted as in Fig.~\ref{Fig:contours}.}           
  \label{Fig:dvdvb}
\end{figure} 

In Sect.~\ref{cmp} it was found that $|\delta v| > \delta v_{\beta}$
in the RH models.  Fig.~\ref{Fig:dvdvb}, upper panel, shows the
velocity spans of density enhancements (identified as having a density
higher than the corresponding smooth value) in the FPP model, and
demonstrates that, after structure has developed, $|\delta v|$ is much
higher than $\delta v_{\beta}$ throughout the whole wind. These high
values essentially stem from the location of the starting points of
the density enhancements, which generally lie \textit{before} the
velocities have reached their post shock values (see
Fig.~\ref{Fig:dvdvb}, middle and lower panels).  By using a $\beta$
velocity law (which in principle corresponds to a stochastic velocity
law with $v_{\rm j}=0$ and $\delta v = \delta v_{\beta}$, see
Fig.~\ref{Fig:vj}) together with the density structure from FPP, we
simulated a RH wind with low velocity spans. Indeed, for the
corresponding intermediate line the equivalent width of the absorption
part was $\sim 35 \%$ lower than that of the original FPP model. 
The strong line, on the other hand, remained saturated, because
the ICM in FPP is not void. So, again, the RH models would in parallel
display de-saturated intermediate lines and saturated strong lines,
were it not for the large velocity spans inside the clumps.

We suggest that the large velocity span inside a shell (clump) is primarily
of kinematic origin, and reflects the formation history of the shell. The
shell propagates outwards through the wind, essentially with a $\beta=1$
velocity law \citep{Owocki88}. Fast gas is decelerated in a strong reverse
shock at the inner rim of the shell. The shell collects ever faster material
on its way out through the wind. This new material collected at higher
speeds resides on the star-facing side, i.e.~at smaller radii, of the slower
material collected before. Thus, a negative velocity gradient develops
inside the shell. The fact that $|\delta v| \gg \delta v_{\beta}$ in FPP
seems to reflect that the shell is formed at small radii, and then advects
outwards maintaining its steep interior velocity gradient\footnote{Actually,
the velocity gradient may further steepen during advection, due to faster
gas trying to overtake slower gas ahead of it; however, this effect is
balanced by pressure forces in the subsonic postshock domain.}. From this
formation in the inner, steeply accelerating wind, velocity spans within the
shells up to (a few) hundred $\rm km\,s^{-1}$, as seen in
Fig.~\ref{Fig:dvdvb}, appear reasonable. 

However, the dynamics of shell formation in hot star winds is very
complex due to the creation and subsequent merging of subshells, as caused
by nonlinear perturbation growth and the related excitation of harmonic
overtones of the perturbation period at the wind base
\citep[see][]{Feldmeier95}. Future work is certainly needed to clarify to
which extent the large velocity spans inside the shells in RH models are a
stable feature (see also Sect.~\ref{future}).

\subsection{3D effects} 
\label{3d}

A shortcoming of our analysis is the assumed symmetry in $\Phi$. The 2D
rather than 3D treatment has in part been motivated by computational reasons
(see Appendix~\ref{rt_code}).  More importantly though, we do not expect our
\textit{qualitative} results to be strongly affected by an extension to 3D.
Within the broken-shell wind model, all wind slices are treated
independently, and distances between clumps increase only in the radial
direction.  Therefore the expected outcome from extending to 3D is a
smoothing effect rather than a reduction or increase in integrated profile
strength (similar to the smoothing introduced by $N_{\Theta}$, see
Sect.~\ref{ang_dep}). Also, we have shown that the main effect from the 
inhomogeneous winds is on the absorption part of the line profiles (see,
e.g., Sect.~\ref{shapes}). The formation of this part is dominated by 
radial photons, especially in the outer wind, because of the dependence 
only on photons released directly from the photosphere.  This implies that
most photons stay within their wind slice, restricting the influence from
any additional `holes' introduced by a broken symmetry in $\Phi$ to the
inner wind. Of course, these expectations hold only within the broken shell
model, because in a real 3D wind the clumps will, for example, have velocity
components also in the tangential directions.

\subsection{Comparison to other studies}
\label{oskow}

To scale the smooth opacity in the formal integral of the non-LTE
atmospheric code PoWR, \citet{Oskinova07} used a porosity formalism in
which both $f_{\rm v}$ and the average distance between clumps enter.
Other assumptions were a void ICM, a smooth $\beta$ velocity field,
and a microturbulent velocity $v_{\rm t} \approx 50 \rm\,km\,s^{-1}$,
the last identified as the velocity dispersion within a
clump. However, a direct comparison between their study and ours is
hampered by the different formalisms used for the spacing of the
clumps. Here we have used the `broken-shell' wind model as a base (see
Sect.~\ref{wind_stoch}), in which each wind slice is treated
independently and the distance between clumps increases only in the
radial direction (clumps preserve their lateral angles).  This gives a
radial number density of clumps, $n_{\rm cl} \propto v^{-1}$, the same
as used by, e.g., \citet{Oskinova06}, when synthesizing X-ray emission
from hot stars.  In \citet{Oskinova07}, on the other hand, the
distance between clumps increases in \textit{all} spatial
directions. In a spherical expansion, this gives a radial number
density of clumps $n_{\rm cl} \propto v^{-1}r^{-2}$, i.e., clumps are
distributed much more sparsely within this model, especially in the
outer wind. Therefore their choice of $L_{\rm 0}=0.2$ is not directly
comparable with $\delta t =0.2$ in our models. The shapes of the
clumps differ between the two models as well; in
\citeauthor{Oskinova07} clumps are assumed to be `cubes', whereas here
the exact shapes of the clumps are determined by the values of the
clumping parameters. Despite these differences, our findings confirm
the qualitative results of \citeauthor{Oskinova07} that the line
profiles become weaker with an increasing distance between clumps as
well as with a decreasing $v_{\rm t}$. These results may be
interpreted on the basis of the effective escape ratio, $\eta$ (see
Eq.~\ref{Eq:heff}).  Both a decrease in $v_{\rm t}$ and an increase in
the distance between clumps mean that the velocity span covered by a
resonance zone becomes smaller when compared to the velocity gap between
two clumps (see Fig.~\ref{Fig:fer}, left panel), leading to higher
probabilities for line photons to escape their resonance zones without
interacting with the wind material.

An important result of this paper is that models that de-saturate
intermediate lines require a non-void ICM to saturate strong
lines. This is confirmed by the \citeauthor{Oskinova07} model, in
which the ICM is void and strong lines indeed do not saturate
\citep{Hamann09}.

\citet{Owocki08} proposed a simplified description of the
non-monotonic velocity field to account for vorosity, i.e., the
velocity gaps between the clumps. Here, the vorosity effect has been
discussed using the quantity $\eta$ (see Sect.~\ref{eta}), and we have
introduced two new parameters to characterize a non-monotonic velocity
field, $\delta v$ and $v_{\rm j}$. The reason for introducing a new
parameterization is that when using a single velocity parameter, we
have not been able to simultaneously meet the constraints from strong,
intermediate, and weak lines as listed in Sect.~\ref{2d}. Tests using
a `velocity clumping factor' $f_{\rm vel}=\delta v / \Delta v$ as
proposed by \citet{Owocki08}, together with a smooth density
structure, have shown that this treatment indeed can reduce the line
strengths of intermediate lines, but that the observational
constraints from strong lines may not be met.  Still, the basic
concept of vorosity holds within our analysis.  For example, one may
phrase the high values of $\delta v$ in the RH models in terms of
insufficient vorosity.

\subsection{Comparison to observations}

\begin{figure}
  \resizebox{\hsize}{!}{\includegraphics[angle=90]{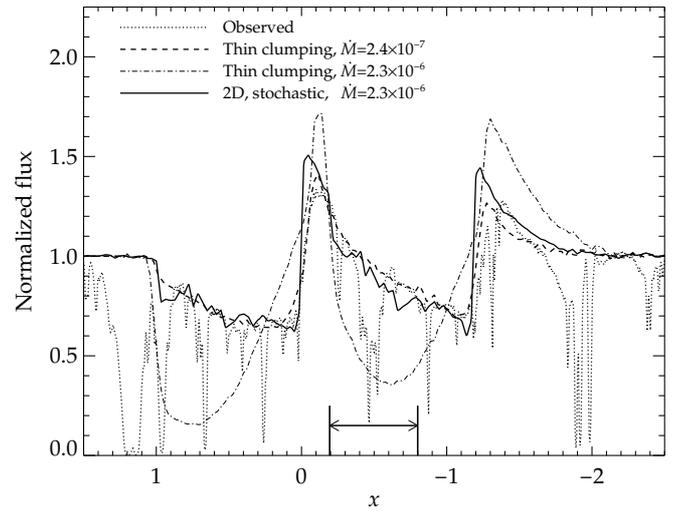}} 
    \caption{Observed FUSE spectra of the PV doublet
    $\lambda\lambda$1118-1128 for the O6 supergiant $\lambda$ Cep 
    \citep{Fullerton06}. The synthetic spectra are calculated for two 1D
    models assuming optically thin clumping (see Sect.~\ref{cmp_obs}) and
    for one 2D stochastic model with parameters as in Table~\ref{Tab:mod}, 
    model Obs1. The models have mass-loss rates $\dot{M}\,[\rm
    M_{\odot}\,yr^{-1}]$ as given in the figure. The zero point
    frequency is shifted to the line center of the $\lambda$1118 component,
    and the two arrows at the bottom of the figure indicate in
    which region the two components overlap.} 
    \label{Fig:cmp_obs}
\end{figure}

We finalize our discussion by performing a first comparison to
observations. The two components of the Phosphorus V
$\lambda\lambda$1118-1128 doublet are rather well separated, and the
singlet treatment used here suffices to model the major part of the
line complex. Nevertheless, the two components overlap within a
certain region (indicated in Fig.~\ref{Fig:cmp_obs}), so when
interpreting the results of this subsection, one should bear in mind
that the overlap is not properly accounted for, but treated as a
simple multiplication of the two profiles.

We used observed FUSE spectra (kindly provided by A. Fullerton) from
HD\,210839 ($\lambda$ Cep), a supergiant of spectral type O6\,I(n)fp.  When
computing synthetic spectra, we first assumed optically thin clumping with a
constant clumping factor $f_{\rm cl}=9$ and a smooth $\beta=1$ velocity
field. $f_{\rm cl}=9$ agrees fairly well with the analysis of
\citet{Puls06}, who derived clumping factors $f_{\rm cl}=6.5$ for $r \approx
1.2 \dots 4.0$ and $f_{\rm cl}=10$ for $r \approx 4.0 \dots 15$, assuming 
an un-clumped outermost wind.\footnote{This stratification has been
found to be prototypical for O-supergiants and was, together with its well
developed PV P Cygni profiles, the major reason for choosing $\lambda$~Cep as
comparison object instead of, e.g., $\zeta$~Pup, which displays a somewhat
unusual run of $f_{\rm cl}$.}
We took the ionization fraction $q=q(r)$ of PV from \citet{Puls08b},
calculated with the unified non-LTE atmosphere code FASTWIND for an O6
supergiant, using the Phosphorus model atom from
\citet{Pauldrach01}. The feedback from optically {\it thin} clumping
was accounted for and X-rays were neglected. This ionization fraction
was then used as input in our MC-1D code when computing the synthetic
spectra.  We assigned a thermal plus a highly supersonic
`microturbulent' velocity $v_{\rm t}=0.05$ (corresponding to 110
km\,s$^{-1}$), as is conventional in this approach.  The mass-loss
rate was derived using the well known relation between $\kappa_0$ and
$\dot{M}$ \citep[e.g.,][]{Puls08}. For atomic and stellar parameters,
we adopted the same values as in \citet{Fullerton06}.

The dashed line in Fig.~\ref{Fig:cmp_obs} represents our fit to the
observed spectrum, assuming optically thin clumping, resulting in a
mass-loss rate $\dot{M}=0.24$, in units of
$10^{-6}\rm\,M_{\odot}\,yr^{-1}$.  \citet{Fullerton06} derived
$\langle q \rangle \dot{M} = 0.23$ for this star. Because our clumped
FASTWIND model predicts an averaged ionization fraction $\langle q
\rangle \approx 0.9$ in the velocity regions utilized by
\citeauthor{Fullerton06}, the two rates are in excellent agreement. On
the other hand, \citet{Repolust04} for HD\,210839 derived
$\dot{M}=6.9$ from $\rm H_{\alpha}$ assuming an unclumped wind,
yielding $\dot{M}_{\rm H_{\alpha}}=2.3$ when accounting for the
reduction implied by our assumed $f_{\rm cl}=9$ ($\dot{M}_{\rm H
  \alpha}=\dot{M}_{\rm H \alpha,sm}f_{\rm cl}^{-1/2}$). This rate is
almost ten times higher than that inferred from PV, and thus results
in PV line profiles that are much too strong (see
Fig.~\ref{Fig:cmp_obs}, dashed-dotted line). That is, to reconcile
the $\rm H_{\alpha}$ and PV rates for HD\,210839 with models that
assume optically {\it thin} clumps also in PV, we would have to
raise the clumping factor to $f_{\rm cl} > 100$.  In addition to this
very high clumping factor, the low rate inferred from the PV lines
conflicts with the theoretical value $\dot{M}=3.2$ provided by the
mass-loss recipe in \citet{Vink00} \citep[using the stellar parameters
  of][]{Repolust04}, and is also strongly disfavored by current
massive star evolutionary models \citep{Hirschi08}.

Next we modeled the PV lines using our MC-2D code together with a
stochastic 2D wind model. The same clumping factor ($f_{\rm cl}=9$)
and ionization fraction (calculated from FASTWIND, see above) were
used.  This time, we assigned $v_{\rm t}=0.005$, i.e., applied no
microturbulence. In previous sections, e.g. \ref{st} and
\ref{shapes}, we showed that stochastic models generally display a
line shape different from smooth models, with a characteristic
absorption dip at the blue edge as well as a dip close to the line
center. Such shapes are not seen in the PV lines in $\lambda$
Cep. Thus, to better resemble the observed line shapes, we used
different values for $\delta t$ and $\fic$ in the inner and outer wind
(the former modification already discussed in Sect.~\ref{outer}) and
let clumping start close to the wind base. Clumping parameters are
given in Table~\ref{Tab:mod}, model Obs1.

As illustrated in Fig.~\ref{Fig:cmp_obs}, the synthetic line profiles
using $\dot{M}=2.3$, as inferred from $\rm H_{\alpha}$, are now at the
observed levels. Because of our insufficient treatment of line
overlap, we gave higher weight to the $\lambda$1118 component when
performing the fitting, but the profile-strength ratio between the
blue and red component was nevertheless reasonably well reproduced
(see also discussion in Sect.~\ref{dens_par}).  However, though the
fit appears quite good, we did not aim for a perfect one, and must
remember the deficits of our modeling technique. For example, while
the early onset of clumping definitely improved the fit (using our
default value, there was a dip close to line center) and might be
considered as additional evidence that clumping starts close to the
wind base, the same effect could in principle be produced by non-LTE
effects close to the photosphere or by varying the underlying $\beta$
velocity law. Such effects will be thoroughly investigated in a
follow-up paper, which will also include a comparison to observations
from many more objects.

Clearly, a consistent modeling of resonance lines (at least of
intermediate strengths) requires the consideration of a much larger
parameter set than if modeling via the standard diagnostics assuming
optically thin clumping, and a reasonable fit to a single observed
line complex can be obtained using a variety of different parameter
combinations.  The analysis of PV lines as done here can therefore, at
present, only be considered as a consistency check for mass-loss rates
derived from other, independent diagnostics, and not as a tool for
directly estimating mass-loss rates. Additional insight might be gained
by exploiting more resonance doublets, due to the different reactions
of profile strengths and shapes on $\kappa_0$. The different slopes of
the equivalent width as a function of $\kappa_0$ in smooth and clumped
models, especially at intermediate line strengths
(Sect.~\ref{dens_par}), may turn out to be decisive. However, because
of, e.g., the additional impact from the ICM density, also this
diagnostics requires additional information from saturated
lines. Taken together, only a consistent analysis using different
diagnostics and wavelength bands, and embedded in a suitable non-LTE
environment, will (hopefully) provide a unique view.
 
\label{cmp_obs}

\section{Summary and future work}
\label{Conclusions}
\subsection{Summary}

Below we summarize our most important findings:

\begin{itemize} 

\item When synthesizing resonance lines in inhomogeneous hot star
  winds, the detailed density structure, the non-monotonic velocity
  field, and the inter-clump medium are all important for the line
  formation.  Adequate models must be able to simultaneously meet
  observational and theoretical constraints from strong, intermediate,
  and weak lines.

\item Resonance lines are basically unaffected by the inhomogeneous
  wind structure in the limit of optically thin clumps, but the clumps
  remain optically thin only for very weak lines.

\item We confirm the basic effects of porosity (stemming from
  optically thick clumps) and vorosity (stemming from velocity gaps
  between the clumps) in the formation of primarily lines of
  intermediate strengths.

\item We point out the importance of a non-void ICM for the
  simultaneous formation of strong and intermediate lines that meet
  observational constraints.

\item Porosity and vorosity are found to be intrinsically coupled and
  of similar importance. To characterize their mutual effect on
  intermediate lines, we have identified a crucial parameter, the
  `effective escape ratio', that describes to which extent photons may
  escape their resonance zones without ever interacting with the wind
  material.
 
\item We confirm previous results that time-dependent,
  radiation-hydrodynamic wind models reproduce observed
  characteristics for strong lines, without applying the highly
  supersonic microturbulence needed in smooth models.

\item A significant profile strength reduction of intermediate lines
  (as compared to smooth models) is for the radiation-hydrodynamic
  models prevented by the large velocity spans of the density
  enhancements, implying that the wind structures predicted by present
  day RH models are not able to reproduce the observed strengths of
  intermediate lines unless invoking a very low mass-loss rate.

\item Provided a non-void ICM and not too large velocity spans inside
  the clumps, 2D \textit{stochastic} wind models saturate strong
  lines, while simultaneously not saturating intermediate lines (that
  are saturated in smooth models).  Using typical volume filling
  factors, $f_{\rm v} \approx 0.25$, the resulting integrated profile
  strength reductions imply that these inhomogeneous models would be
  compatible with mass loss rates roughly a factor of ten higher than
  those derived from resonance lines using smooth models.

\item A first comparison to observations was made for the O6
  supergiant $\lambda$ Cep. It was found that, indeed, the line
  profiles of PV based on a 2D stochastic wind model, accounting for a
  detailed density structure and a non-monotonic velocity field,
  reproduced the observations with a mass-loss rate almost ten times
  higher than the rate derived from the same lines, but with a model
  that used the optically thin clumping approach.  This alleviated the
  discrepancies between theoretical predictions, evolutionary
  constraints, and previous mass-loss rates based on winds assumed
  either to be smooth or to have optically thin clumps.

\end{itemize}

\subsection{Future work} 
\label{future} We have investigated general properties of resonance line
formation in inhomogeneous 2D wind models with non-monotonic velocity
fields.  To perform a detailed and quantitative comparison to
observations, and derive mass-loss rates, simplified approaches need
to be developed and incorporated into non-LTE models to obtain
reliable occupation numbers.  Extending our Monte-Carlo radiative
transfer code to include line overlap effects in doublets is critical
for more quantitative applications, and an extension to 3D is also
necessary. Further applications involve synthesizing emission lines,
for example to test the optically thin clumping limit both in the
parameter range where this is thought to be appropriate (e.g., for
O-/early B-stars), and in other more complicated situations.  Indeed,
the present generation of line-blanketed model atmospheres does not
seem to be able to reproduce $\rm H_{\alpha}$ line profiles from
A-supergiants, which are observed as P-Cygni profiles with
\textit{non-saturated} troughs, whereas the simulations (assuming
optically thin clumping) result in saturated troughs (R.-P. Kudritzki,
private communication). Since $\rm H_{\alpha}$ is a quasi-resonance
line and not a recombination line in these cooler winds
\citep[e.g.,][]{Kudritzki00}, this behavior might be explained by the
presence of optically thick clumps.

Finally, it needs to be clarified if the large velocity span inside
clumps generated in RH models is independent of additional physics
that is not, or only approximately, accounted for in present
simulations (such as more-D effects and/or various exciting
mechanisms). If the large velocity span is a stable feature, one might
come to the (rather unfortunate) conclusion that either the observed
clumping features are not, or only weakly, related to the line-driven
instability, or the discrepancies between observed and synthetic flux
distribution (from the X-ray to the radio regime) might involve
processes different from the present paradigm of wind clumping.
 
\begin{acknowledgements}
{We would like to acknowledge our anonymous referee and A. Fullerton
  for useful comments and suggestions on the first version of this
  manuscript. Many thanks to A. Fullerton also for providing us with
  reduced PV spectra for his O-star sample, and W.-R. Hamann for
  suggesting the term `velocity span' for the parameter $\delta
  v$. K. Lind is also thanked for a careful reading of the manuscript.
  J.O.S gratefully acknowledges a grant from the International
  Max-Planck Research School of Astrophysics (IMPRS), Garching.}

\end{acknowledgements}

\bibliographystyle{aa}

\appendix

\section{The Monte-Carlo transfer code}
\label{rt_code}

\subsection{The code}

Here we describe our Monte-Carlo radiative transfer code (MC-2D) in some
detail. For an overview of basic assumptions, see Sect.~\ref{rt} in the main
paper. For testing purposes, versions to treat spherically symmetric winds, 
either in the Sobolev approximation (MCS-1D) or exactly (MC-1D), have
been developed as well. 

\begin{figure}
  \centering
\resizebox{\hsize}{!}{\includegraphics[angle=90]{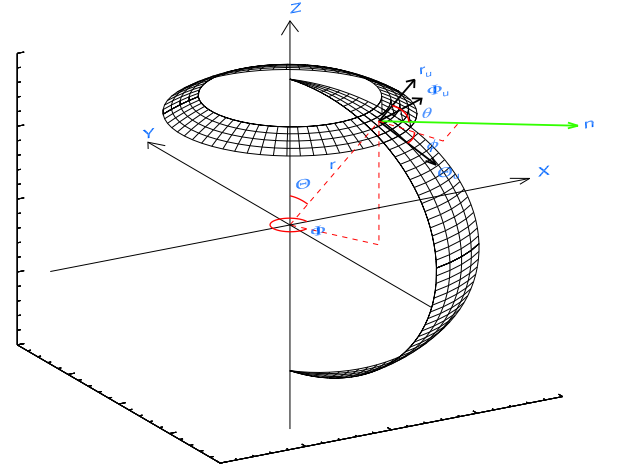}}              
                 \caption{Illustration of the coordinate system, see text. 
                   \textit{A color version of this figure is available in the web version.}}
                 \label{Fig:coord}
\end{figure}

\paragraph{Geometry.}
For wind models in which the spherical symmetry is broken, we can no
longer restrict photon trajectories to rays with constant impact parameters
(see below). Moreover, the observed spectrum will depend on the observer's
placement relative to the star. Fig.~\ref{Fig:coord} illustrates the
geometry in use, a standard right-handed spherical system ($r,\Theta,\Phi$)
defined relative to a Cartesian set ($X,Y,Z$) (transformations between the
two may be found in any standard mathematical handbook).  At each coordinate
point we also construct a local coordinate system using the local unit
vectors $(r_{\rm u},\Theta_{\rm u},\Phi_{\rm u})$, which for a photon
propagating in direction $n_{\rm u}$ is related to the \textit{radiation
coordinates} $(\theta,\phi)$ (see Fig.~\ref{Fig:coord}) via
\begin{equation} 
  \cos \theta \equiv \mu = r_{\rm u} \cdot n_{\rm u},
  \label{Eq:setb}
\end{equation}
\begin{equation}
  \sin \phi \sin \theta = \Phi_{\rm u} \cdot n_{\rm u} = \frac{Z_{\rm u} 
    \times r_{\rm u}}{|Z_{\rm u} \times r_{\rm u}|} \cdot n_{\rm u},
\end{equation}
\begin{equation}
  \cos \phi \sin \theta = \Theta_{\rm u} \cdot n_{\rm u} = [\Phi_{\rm u} \times r_{\rm u}] \cdot n_{\rm u}.
\end{equation}
The radiation coordinates are defined on the intervals 
$\theta = 0 \dots \pi$ and $\phi = 0 \dots 2 \pi$, but due to the
symmetry in $\Phi$, only the range $\phi = 0 \dots \pi$ needs to be considered 
(see \citeauthor{Busche00} 2000). Also, for this symmetry, 
the direction cosines of $n_{\rm u}$ simplify to 
\begin{equation} 
  n_{\rm x} = \mu \sin \Theta + \sqrt{1-\mu^2} \cos \phi \cos \Theta, 
\end{equation}
\begin{equation}
  n_{\rm y} =  \sqrt{1-\mu^2} \sin \phi,
\end{equation}
\begin{equation}
  n_{\rm z} = \mu \cos \Theta - \sqrt{1-\mu^2} \cos \phi \sin \Theta.
  \label{Eq:sete}
\end{equation}
Eqs.~\ref{Eq:setb}-\ref{Eq:sete} are used to update the 
physical position ($r,\Theta$) of the photon and the 
local values of the radiation coordinates ($\theta,\phi$).
By tracking the photon on a radial mesh, both the physical 
and radiation coordinates can be updated exactly. 
Interpolations are necessary only when a photon is scattered or when it 
crosses a $\Theta$-boundary to another wind slice. 
Essentially the same coordinate system is used by, e.g., 
\citet{Busche00}. 
We collect escaped photons according to their 
$\Theta$-angles at `infinity'\footnote{The full 3D problem would require binning in $\Phi$ as well,
which in turn would require a large increase in the number of simulated photons.}, 
and bin them using the same $N_{\Theta}$ bins as in the 
underlying wind model (see Sect.~\ref{wind}).   

For spherically symmetric wind models, we adhere 
to the customary $(p,z)$ spatial coordinate system with $p$ being the 
impact parameter and $z$ the direction toward the observer. 
Each time a photon is scattered and its direction determined, a new impact parameter is 
computed from the relation $p=r\sqrt(1-\mu^2)$, appreciating 
that all points on a surface of constant radius can be treated equally in this geometry.

\paragraph{Releasing photons.} We release photons from the lower boundary
uniformly in $\phi$ and with a distribution function $\propto \mu d\mu$ in $\mu$ 
\citep[e.g.,][]{Lucy83}. The angular coordinate
$\Theta$ is selected so that photons are uniformly distributed over the surface area 
$dA=\sin \Theta d\Theta d\Phi$.  

\paragraph{Absorption.} The probability of photon absorption is 
$\propto e^{-\tau}d\tau$, hence the optical depth $\tau$ 
the photon travels before absorption can be selected according to 
$\tau = - \ln {R_{\rm 1}}$, where $R_{\rm 1}$ is a random number between 0 and 1. 
The position for absorption in the wind may then be determined by inverting the line
optical depth integral along the photon path
\begin{equation} 
	\label{Eq:tau}
  \tau_{\nu} = \int \chi_{\nu} ds,  
\end{equation} 
\noindent with the frequency-dependent opacity
\begin{equation}
  \chi_{\nu} = \kappa_{\rm L}\rho \phi_{\nu}, 
\end{equation}
\noindent with $\phi_{\nu}$ the absorption profile, $\kappa_{\rm L}$ the 
frequency integrated mass absorption coefficient, and $\rho$ the mass
density.  All dependencies on spatial location are for simplicity suppressed
here and in the following.  For the opacity we use the parameterization from
\citet{Hamann81} and POF,
\begin{equation}
  \label{Eq:kappa0}
  \kappa_{\rm L}\lambda \rho = \frac{4\pi R_{\star}\vinf^2}{\dot{M}}\kappa_{\rm 0}\rho q,
\end{equation}
\noindent where $\lambda$ is the wavelength of the considered transition, 
$\kappa_{\rm 0}$ is a `line-strength' parameter taken
to be constant, $\dot{M}$ the radially and laterally averaged
mass-loss rate, and $q=q(r,\Theta)$ the fraction of the considered
element that resides in the investigated ionic stage. Default here is
$q=1$, but effects from other ionization structures are discussed in
Sect.~\ref{outer}.  $\kappa_{\rm 0}$ is proportional to the product of
mass-loss rate and abundance of the considered ion, and, for a smooth
wind, $\kappa_{\rm 0}=1$ and $\kappa_{\rm 0}=100$ give a typical
medium and strong line, respectively. 
The parameterization as defined in Eq.~\ref{Eq:kappa0} has the advantage
that for smooth winds the radial optical depth in the Sobolev approximation collapses to
\begin{equation}
  \tau_{\rm Sob} = \frac{\kappa_{\rm 0}}{r^2 v {\rm d}v/{\rm d}r}\, q,
\end{equation}
when $v$ and $r$ are expressed in normalized units. The corresponding
expression for clumpy winds is
provided in Eq.~\ref{Eq:tau_s}.
The absorption profile is assumed to be a Gaussian with a Doppler width
$v_{\rm t}$ that contains the contributions from thermal and (if present)
`microturbulent' velocities. To solve Eq.~\ref{Eq:tau}, we adopt the
dimensionless frequency $x$ with the terminal velocity of a smooth
outflow as the reference speed,
\begin{equation}
x=\frac{\nu-\nu_{\rm 0}}{\nu_{\rm 0}} \frac{c}{\vinf},
\label{Eq:x}
\end{equation}
\noindent and transform to the co-moving frame (hereafter CMF).
$\nu_{\rm 0}$ is the rest-frame frequency of the line center and $c$
the speed of light.  We now assume that between two grid points the
variation of the factor $\kappa_{\rm L}\rho/|Q|$ (see below) is small
and may be replaced by an average value.  The optical depth $\Delta
\tau_{\nu}$ between two subsequent spatial points $(r,\Theta)$ then
becomes
\begin{equation} 
  \label{Eq:dtau}
  \Delta \tau_{\nu} = |\frac{\lambda R_\star}{\vinf} \, \frac{ \kappa_{\rm L}\rho}{Q} 
  \times \frac{-\Delta \rm erf[\it x_{\rm cmf}/v_{\rm t}]}{2}|,
\end{equation}
\noindent where $\Delta \rm erf$ is the difference of the error-function
between the points, $x_{\rm cmf}$ the dimensionless CMF frequency, and $v_t$
is calculated in units of $\vinf$. $Q~\equiv~n_{\rm u}~\cdot~\nabla~(n_{\rm
u}~\cdot~\vec{v}) $ is the local directional derivative of the velocity in
direction $n_{\rm u}$, with velocities measured in units of $\vinf$ and
radii in units of $R_\star$. By interpolating to the border whenever a
photon crosses a $\Theta$ boundary, we \textit{locally} recover the
spherically symmetric expression
\begin{equation}
  Q = \frac{\partial v}{\partial r}\mu^2+\frac{v}{r}(1-\mu^2).
\end{equation}
For spherically symmetric winds, we have written a second implementation 
that allows for line transfer using the Sobolev approximation.  With this
method each resonance zone is approximated by a point and the line only
collects optical depth at atmospheric locations where the observer's frame
frequency $x_{\rm obs}$ has been Doppler shifted to coincide with the CMF
frequency for the line center.  The condition for interaction thus is
$x_{\rm obs}=\mu v$ and the last factor in Eq.~\ref{Eq:dtau} collapses to
unity when calculating the Sobolev optical depth.  The Sobolev approach can
be expected a reasonable approximation when the variation of the factor
$\kappa_{\rm L}\rho/|Q|$ is small within the whole resonance zone
contributing to the optical depth in Eq.~\ref{Eq:dtau}, i.e., small on
length scales at least a few times the Sobolev length $L \equiv v_{\rm
t}/|Q|$. However, also in the Sobolev approximation more than one resonance
point may be identified in a wind with a non-monotonic velocity field.

\paragraph{Re-emission.} We assume complete redistribution and isotropic
re-emission in the CMF, allowing for a multitude of scattering events within
one resonance zone.  When the Sobolev approximation is applied, re-emission 
is assumed to be coherent in the CMF and for the angular re-distribution we
then use the corresponding escape probabilities \citep{Castor70}, corrected
for a treatment of negative velocity gradients (\citeauthor{Rybicki78}
1978;\,POF). In this case, there is only one effective scattering event
inside the localized resonance zone. 

After the photon has been re-emitted at some atmospheric location, the
procedure runs again and searches for another absorption. 

\subsection{Radiative transfer code tests}
\label{rt_tests}

\begin{figure}
    \resizebox{\hsize}{!}{\includegraphics[angle=90]{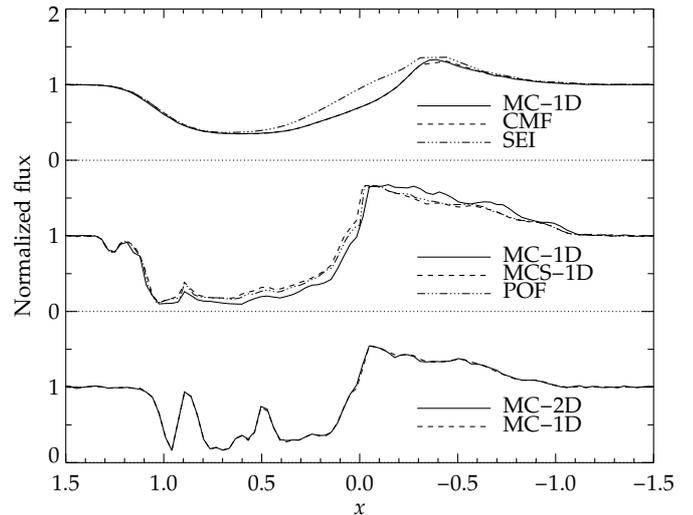}} 
	     \caption{Synthetic line profiles for spherically symmetric
	     models, calculated with the labeled methods. Profiles are shown
	     for a smooth model with $\kappa_0=1.0$ and $v_{\rm t}=0.2$
	     (upper) and for two POF snapshots with $\kappa_0=100$ (middle)
	     and $\kappa_0=5.0$ (lower) and $v_{\rm t}=0.005$. The 2D
	     profile is for an observer at the equator.  $x$ is the
	     normalized observer's frame frequency (see Eq.~\ref{Eq:x}), and
	     the ordinate displays the emergent flux normalized to the
	     continuum flux.} 
	     \label{Fig:1d_prof}
\end{figure} 

In this subsection we describe some of the verification tests of our MC
radiative transfer code that we have made. The MC-1D version was first
applied on spherically symmetric winds, comparing profiles from smooth,
stationary winds to profiles calculated using the well-established CMF
(cf.~\citeauthor{Mihalas75} 1975; \citeauthor{Hamann81} 1981) and SEI
methods, and profiles from time-dependent RH winds to profiles calculated
using the Sobolev method developed in POF. Thereafter we applied the MC-2D
version on models in which all lateral slices had the same radial structure,
comparing the results to the MC-1D version.
     
First we calculated line profiles for smooth, 1D winds. We have verified
that for low\footnote{For a typical terminal velocity value $\vinf=2000 \
\rm km \ s^{-1}$, $v_{\rm t}=0.005$ corresponds to $10 \ \rm km \ s^{-1}$
and $v_{\rm t}=0.2$ to $400\ \rm km \ s^{-1}$.} values of $v_{\rm t}$,
profiles from all the methods described above agree perfectly, whereas for
higher values the MC-1D and CMF give identical results but the SEI deviates 
significantly, especially for a medium-strong line (see
Fig.~\ref{Fig:1d_prof}, upper panel).  This is due to the hybrid nature of
the SEI technique, which approximates the source function with its local
Sobolev value but carries out the exact formal integral. Because of this,
the method does not account for the increasing amount of photons close to
line center that are backscattered into the photosphere when the resonance
zone grows and overlaps with the lower boundary.\footnote{Remember that
neither the SEI nor the CMF, as formulated here, include a transition to the
photosphere, but treat the lower boundary as sharp with a minimum velocity
$v_{\rm min}$.} Consequently the re-emitted flux in this region is higher
when calculated via the SEI than when calculated via the CMF or MC methods. 
These discrepancies between the CMF and SEI are quite well documented and
discussed \citep[e.g.,][]{Hamann81,Lamers87}, however we still emphasize
that one should exercise caution when applying the SEI method with high 
microturbulence on wind resonance lines. Especially today, when increased
computer-power enables us to compute fast solutions using both methods, the
CMF is preferable.  

Next we calculated line profiles for structured, 1D winds.  Profiles
computed with all three methods agreed for weak and intermediate lines.  For
strong lines, the agreement between MCS-1D and the method from POF, which
uses a Sobolev source function accounting for multiple-resonance points, was
satisfactory.  However, minor discrepancies between Sobolev and non-Sobolev
treatments occurred for the strong line also when no microturbulent velocity
was applied (see Fig.~\ref{Fig:1d_prof}), as opposed to the smooth case.

Finally we performed a simple test of our MC-2D code by applying it on
models in which all lateral slices had the same radial structure, i.e., the
wind was still spherically symmetric and all observers ought to see the same
spectrum. We confirmed that indeed so was the case, both for smooth and
structured models (in Fig.~\ref{Fig:1d_prof} the latter case is
demonstrated).  


\section{The effective escape ratio}
\label{app_eta}
We define the ratio of the velocity gap $\Delta v$ between two clumps (see
Fig.~\ref{Fig:fer} in the main paper) and the thermal velocity $v_{\rm t}$
as 
\begin{equation} 
  \eta \equiv \frac{\Delta v}{v_t}.
\end{equation}
In the following, we derive an expression for $\eta$, for the wind geometry
used throughout this paper.  If $\Delta v_{\rm tot} = \Delta v + |\delta v|$
is the velocity difference between two clump \textit{centers}, we may write
(omitting the absolute value signs here and in the following)
\begin{equation}
  \Delta v = \Delta v_{\rm tot} - \delta v = \frac{\Delta v_{\rm tot} }
  {\Delta v_{\rm tot,\beta}} \Delta v_{\rm tot,\beta}
  -\frac{\delta v }{\delta v_{\beta}} \delta v_{\beta},
\end{equation}
where we have normalized the arbitrary velocity intervals to the
corresponding $\beta$ intervals. $\beta$ suffixes are used to denote
parameters of a smooth velocity law.  For notational simplicity we write
\begin{equation}
  \xi_1 = \frac{\Delta v_{\rm tot} }{\Delta v_{\rm tot,\beta}},  \qquad
  \xi_2 = \frac{\delta v }{\delta v_{\beta}}. 
\end{equation}
Assuming radial photons, $\Delta v$ may be approximated by 
\begin{equation}
  \Delta v \approx \frac{\partial v_{\beta}}{\partial r} \Delta r_{\rm tot,\beta}
  (\xi_{\rm 1}  - \xi_{\rm 2} \frac{\delta r_{\beta}}{\Delta r_{tot,\beta}} ),
  \label{Eq:Dv}
\end{equation}
with the notations of $r$ following those of $v$. The volume filling factor
for the geometry in use is 
\begin{equation}
f_{\rm v} \equiv \frac{V_{\rm cl}}{V_{\rm tot}} \approx \frac{r_{\rm 1}^2 \delta r}{r_{\rm 2}^2 \Delta r_{\rm tot}}
\label{Eq:fv}
\end{equation}
with $V_{\rm cl}$ the volume of the clump, $V_{\rm tot}$ the total volume,
and $r_{\rm 1}~\approx~r_{\rm 2}$ the radial points associated with the
beginning of the clump and the ICM.  Using Eq.~ \ref{Eq:fv} and $\Delta
r_{\rm tot} = v_{\beta} \delta t$ (see Sect.~\ref{wind_stoch}), we obtain 
\begin{equation}
  \Delta v \approx \frac{\partial v_{\beta}}{\partial r} v_{\beta} \delta t
  ( \xi_1 - \xi_2 f_v ), 
\end{equation}  
and for $\eta$, using the radial Sobolev length of a smooth flow $L_{\rm
r}~=~v_{\rm t}/(\partial v_{\beta}/\partial r)$,
 \begin{equation}
  \eta \approx \frac{v_{\beta} \delta t
  ( \xi_1 - \xi_2 f_v )}{L_{\rm r}}.
\end{equation}  
In our models $\xi_1$ is not given explicitly, but is on the order of unity,
because we distribute clumps according to the underlying smooth $\beta =1$
velocity law. Thus we approximate 
\begin{equation} 
  \eta \approx \frac{v_{\beta} \delta t
  ( 1 - \xi_2 f_v )}{L_{\rm r}}.
 \label{Eq:fesc}
\end{equation}
 
We notice that the porosity length $h$ as defined by \citet{Owocki04} is $h
= l/f_{\rm v}$, where $l$ is the length associated with the clump.  For the
geometry used here this becomes $h \approx \delta r/f_{\rm v} \approx
v_{\beta} \delta t$.  Hence, using $\xi_2=1$ for a smooth velocity field,
$\eta$ represents the porosity length corrected for the finite size of the
clump, and divided by the radial Sobolev length.  

\end{document}